\documentclass[
 journal=pasa,
 manuscript=article-type,
 year=2025,
 volume=37,
]{cup-journal}

\usepackage[nopatch]{microtype}
\usepackage{booktabs}
\usepackage{tabularx} 
\usepackage{graphicx}
\usepackage{amsmath,amssymb,bm,enumitem}
\usepackage{multirow}
\usepackage{caption, lipsum }
\usepackage{mathtools}
\usepackage{lineno}
\usepackage{stfloats, placeins}
\usepackage{xcolor}
\usepackage{xurl}
\usepackage{hyperref}

\newcommand{\red}[1]{#1}

\newcommand{\kms}{\,\textrm{km}\,\textrm{s}^{-1}}
\newcommand{\kmsmpc}{$\,\textrm{km}\,\textrm{s}^{-1}\, \textrm{Mpc}^{-1}$ }
\newcommand{\arcsec}{arcsec}
\newcommand{\hikpc}{h^{-1}\textrm{kpc}}

\newcommand{\blist}{\begin{itemize}}
\newcommand{\elist}{\end{itemize}}

\title{The DESI DR1 Peculiar Velocity Survey: Fundamental Plane Catalogue}

\author{C.~E.~Ross,\textsuperscript{1}
C.~Howlett,\textsuperscript{1}
J.~R.~Lucey,\textsuperscript{2}
K.~Said,\textsuperscript{1}
T.~M.~Davis,\textsuperscript{1}
J.~Aguilar,\textsuperscript{3}
S.~Ahlen,\textsuperscript{4}
A.~J.~Amsellem,\textsuperscript{5}
J.~Bautista,\textsuperscript{6}
S.~BenZvi,\textsuperscript{7}
D.~Bianchi,\textsuperscript{8,9}
C.~Blake,\textsuperscript{10}
D.~Brooks,\textsuperscript{11}
A.~Carr,\textsuperscript{12}
T.~Claybaugh,\textsuperscript{3}
A.~Cuceu,\textsuperscript{3}
A.~de la Macorra,\textsuperscript{13}
B.~Dey,\textsuperscript{14,15}
P.~Doel,\textsuperscript{11}
K.~Douglass,\textsuperscript{7}
S.~Ferraro,\textsuperscript{3,16}
A.~Font-Ribera,\textsuperscript{17}
J.~E.~Forero-Romero,\textsuperscript{18,19}
E.~Gaztañaga,\textsuperscript{20,21,22}
S.~Gontcho A Gontcho,\textsuperscript{3,23}
G.~Gutierrez,\textsuperscript{24}
J.~Guy,\textsuperscript{3}
K.~Honscheid,\textsuperscript{25,26,27}
D.~Huterer,\textsuperscript{28,29}
M.~Ishak,\textsuperscript{30}
R.~Joyce,\textsuperscript{31}
A.~G.~Kim,\textsuperscript{3}
A.~Kremin,\textsuperscript{3}
O.~Lahav,\textsuperscript{11}
C.~Lamman,\textsuperscript{27}
M.~Landriau,\textsuperscript{3}
L.~Le~Guillou,\textsuperscript{32}
A.~Leauthaud,\textsuperscript{33,34}
M.~E.~Levi,\textsuperscript{3}
P.~Martini,\textsuperscript{25,35,27}
A.~Meisner,\textsuperscript{31}
R.~Miquel,\textsuperscript{36,17}
J.~Moustakas,\textsuperscript{37}
A.~Muñoz-Gutiérrez,\textsuperscript{13}
S.~Nadathur,\textsuperscript{21}
N.~Palanque-Delabrouille,\textsuperscript{38,3}
W.~J.~Percival,\textsuperscript{39,40,41}
C.~Poppett,\textsuperscript{3,42,16}
F.~Prada,\textsuperscript{43}
I.~Pérez-Ràfols,\textsuperscript{44}
F.~Qin,\textsuperscript{6}
G.~Rossi,\textsuperscript{45}
E.~Sanchez,\textsuperscript{46}
D.~Schlegel,\textsuperscript{3}
M.~Schubnell,\textsuperscript{28,29}
D.~Sprayberry,\textsuperscript{31}
G.~Tarlé,\textsuperscript{29}
R.~J.~Turner,\textsuperscript{10}
B.~A.~Weaver,\textsuperscript{31}
R.~Zhou,\textsuperscript{3}
and H.~Zou\textsuperscript{47}}

\affiliation{Affiliations are listed at the end of the paper.}

\email[C. E. Ross]{c.ross1@uq.net.au}


\begin{document}

\makeatletter
\renewcommand{\@evenhead}{\sffamily\thepage\hfill C.~Ross\space et al.}
\makeatother

\begin{abstract}
Measurements of peculiar velocities in the local Universe are a powerful tool to study the nature of dark energy at low ($z < 0.1$) redshifts. Here we present the largest single set of $z<0.1$ peculiar velocity measurements to date, obtained using the Fundamental Plane (FP) of galaxies in the first data release (DR1) of the Dark Energy Spectroscopic Instrument (DESI). We describe the photometric and spectroscopic selection criteria used to define the sample, as well as extensive quality control checks on the photometry and velocity dispersion measurements. Additionally, we perform detailed systematics checks for the many analysis parameters in our pipeline. Our DESI DR1 catalogue contains FP-based distances and peculiar velocities for $98,292$ unique early-type galaxies, increasing the total number of $z < 0.1$ FP distances ever measured by a factor of $\sim2$. We achieve a precision of $26\%$ random error in our distance measurements which is comparable to previous surveys. A series of companion DESI papers use the distances and peculiar velocities presented in this paper to measure cosmological parameters.
\end{abstract}

\section{Introduction}

It is an exciting time for cosmology, with ever more precise measurements of the expansion history of the Universe revealing cracks in our standard model. Measurements of the local expansion rate of the Universe $H_0$ from direct distance-ladder measurements disagree significantly with those from global fits to the cosmic microwave background and other early-universe measurements \citep[e.g.][]{Planck_Colab_2020_VI,Riess_2022}. Meanwhile, measurements of Type Ia supernovae and baryon acoustic oscillations both hint that dark energy may have a  time-varying equation of state \citep{DESSN_2024, DESIBAO_2024, DESIBAO_2025}.

One method to study these questions at low redshift is using peculiar velocity (PV) catalogues, which use distance indicators to break the degeneracy between the observed and cosmological redshifts of galaxies in our local universe. Doing so allows us to measure the local `peculiar' velocities of these galaxies, and provides the necessary information to study the nature of dark energy and dark matter at low redshifts where other methods are complicated by large scale structure, galaxy bias and the cosmic variance limit. Due to this, PV surveys are commonly used to test our cosmological models through measurements of the local expansion rate, $H_0$, and the structure growth rate, $f\sigma_8$.  

Two of the most common distance indicators used to measure peculiar velocities are the Fundamental Plane relation \citep[FP,][]{Djorgovski_Davis_1987, Dressler_1987} and the Tully-Fisher relation \citep[TF,][]{Tully_Fisher_1977}. The FP relation is a 3-dimensional empirical relationship for early-type galaxies between central velocity dispersion, mean surface brightness and angular effective radius. As the central velocity dispersion and mean surface brightness are distance-independent observables, they can be used to infer the physical effective radius of each galaxy. Comparing this to the distance-dependent angular effective radius allows the Fundamental Plane to be used as a distance indicator. The TF relation is the FP analogue for late-type galaxies and uses rotational velocity as a distance-independent observable to predict absolute magnitude. By comparing this absolute magnitude to the observed apparent magnitude, one can extract distance estimates for the galaxies.

Historically, large PV surveys have used only one of these indicators to obtain distance measurements --- e.g.\ the SMAC sample \citep{Hudson_2001}, the EFAR sample \citep{Colless_2001}, the ENEARc Cluster sample \citep{Bernardi_2002},  the 6dF Galaxy Survey \citep{Springob_2014} and Sloan Digital Sky Survey \citep[SDSS,][]{Howlett_2022} using FP; and the SFI++ survey \citep{Springob_2007}, the Two Micron All Sky Survey \citep{Hong_2019} and the Arecibo Legacy Fast ALFA Survey \citep{Kourkchi_2020} using TF. The largest currently available PV catalogue is Cosmicflows-4 \citep{cosmicflows-4} which combines almost all previous peculiar velocity surveys into a single dataset. The compilation includes measurements for $\sim 55,000$ galaxies across 8 different tracers and 38 different calibrations, with the largest contributions coming from \citet[$\sim 35,000$ FP galaxies,][]{Howlett_2022}  and \citet[$\sim 10,000$ TF galaxies,][]{Kourkchi_2020}. 

The Cosmicflows-4 catalogue has been a valuable resource for tests of $\Lambda$CDM.  Recent results include measurements of the local expansion rate, $H_0 = 74.6\pm 0.8$(stat) $\pm 2.0$(sys) \kmsmpc \hfill \citep{cosmicflows-4}, and growth rate of structure $f\sigma_8 = 0.38 \pm0.04$(stat) (grouped) , $f\sigma_8 = 0.36 \pm0.05$(stat) (ungrouped), $f\sigma_8 = 0.30 \pm0.06$(stat) \citep[supernovae,][]{Courtois_2023}, $f\sigma_8 = 0.35\pm 0.03$ (stat) \citep[TF,][]{Boubel_2024} \red{and $f\sigma_8 = 0.424 \pm0.016$ (stat) \citep[combined TF, FP and supernovae;][]{Stiskalek_2025}.} PVs have also seen substantial application to cosmography, notably \citet{Courtois_2013,Tully_2014,Courtois_2023} \red{and \citet{Giani_2024}} who use them to make detailed maps of our local Universe. Additionally, PVs have been used to measure bulk flow and low-order velocity measurements \red{\citep[and references therein]{Rubin_1976, Hudson_1999, Scrimgeour_2016,Qin_2018, Qin_2019a, Boruah_2020, Qin_2021a, Qin_2021b,   Courtois_2023,  Lopes_2024},} some of which have also noted discrepancies with $\Lambda$CDM \citep{Watkins_2009, Feldman_2010, Peery_2018, Howlett_2022, Whitford_2023, Watkins_2023}.

Despite relatively small statistical uncertainties in the CosmicFlows - 4 catalogue, \red{as can be seen from the above constraints, systematic errors currently dominate $H_0$ constraints, and systematic variations (of up to $1.6\sigma$) between $f\sigma_8$ measurements arise depending on how galaxies are grouped and which tracers/surveys are used. This highlights the difficulty in combining datasets from independent surveys. Conducting a PV survey as one the secondary target programs, the Dark Energy Spectroscopic Instrument \citep[DESI,][]{DESI_collab_2016a} collaboration aims to simplify this problem. The DESI PV survey will create a large homogeneous PV catalogue using both FP and TF distance indicators to obtain distances and peculiar velocities for $\sim186,000$ galaxies over the 5-years of observing operations \citep{DESI_PV_target_selection}. Calibrated in tandem, these two tracers use the same instrument and photometry reducing the possible sources for discrepancies and providing the ability to test for tracer specific biases during the calibration process. The underlying DESI redshift catalogue also provides a single sample with which to robustly identify groups for cross-calibration and zero-pointing.}

The Dark Energy Spectroscopic Instrument (DESI) is a ground based spectroscopic survey conducted on the 4-metre Mayall Telescope at Kitt Peak National Observatory, Arizona, USA, that during its 5 years of operations aims to target 40 million galaxies and quasars across its 14,000 deg$^2$ footprint \citep{DESI_collab_2022_Inst}. This unprecedented survey speed is facilitated by the instrument's $~3^\circ$ field of view and the 5,000 robotically positioned fibres on the instrument focal plane, allowing for simultaneous observation of up to 5,000 individual targets and reconfiguration interruptions of under 3-minutes between exposures \citep{DESI_collab_2016b,DESI_fiberSystem_Poppett_2024,DESI_Corrector_Miller_2023,DESI_SurveyOps_Schlafly_2023}. 

During the first year of operations (DR1), DESI measured spectra for more than 18 million targets across just over 9000 deg$^2$ of sky coverage \citep{DESI_collab_DR1}. This large dataset has allowed the collaboration to use the BAO to conduct the highest precession measurement of the Universe's expansion history to date \citep{DESI_DR1_BAO_cosmology}. \red{Compared to the expected full survey from \cite{DESI_PV_target_selection}, the DR1 FP sample covers $\sim60\%$ of the expected sky area, $\sim40\%$ of the target sample and $\sim70\%$ of the projected successful FP sample. We note however that these comparisons are only rough --- the current sky area is largely incomplete, consisting of mostly single-passes, while the number of successful FP galaxies is higher than expected due to differences in our spectroscopic selection function between this work and that assumed in \cite{DESI_PV_target_selection}. An alternative comparison can be made by considering that the FP component of the DESI PV survey progresses at roughly the same rate as the full bright-time survey which is currently at just over $41\%$ complete \citep[Section 5,][]{DESI_collab_DR1}. This puts the FP survey on track to exceed the projected $133,000$ FP galaxies by $\sim 100,000$ galaxies if the current success rate continues.}

This work, and accompanying parallel work conducted by \citet{DESI_DR1_TF} introduce the DESI DR1 peculiar velocity catalogues. In \citeauthor{DESI_DR1_TF}, the emphasis is on the Tully-Fisher sample, while this work is on the Fundamental Plane sample. These two datasets are combined in \citet{DESI_DR1_PV_zp} and are used to measure the local expansion rate $H_0$ and the systematic uncertainties relating to the zero-pointing are investigated. Additionally, \citet{DESI_DR1_PV_Pantheon_cross_corr}, \citet{DESI_DR1_PV_dens_vel_corr}, \citet{DESI_DR1_PV_maxlike} and \citet{DESI_DR1_PV_power_spec} use a variety of methods to obtain measurements of structure formation rate, $f\sigma_8$, from the combined sample. Finally, \citet{DESI_DR1_PV_mocks} have created a suite of mock catalogues to accompany the DR1 PV datasets from the AbucusSummit N-body simulation suite, which have been used to test our analysis pipelines for the $H_0$ and $f\sigma_8$ measurements.

This paper presents the DESI {\em Fundamental Plane} PV catalogue for the first major DESI data release DR1, extending the work conducted on the early data release described in \citet{Said_EDR}. As part of this analysis, we investigate potential sources of systematic errors with a particular focus on our velocity dispersion measurements and the cuts and corrections applied during the calibration process. The paper is organised as follows: Section \ref{selection:primary} introduces our dataset and selection criteria. In Section \ref{sec:mag_offset_test} we re-evaluate the magnitude offset noticed by \citet{Said_EDR} between the Dark Energy Camera Legacy Survey (DECaLS) and the BASS/MzLS surveys in the DESI Legacy Survey DR9 catalogue. In Section \ref{section:vdisp_measurments} we describe the process through which we obtain our velocity dispersion measurements as well as test for systematics through internal and external consistency checks. We introduce the Fundamental Plane in Section \ref{section:fit_methodology}, describing the derivation of our parameters as well as our fitting methodology and zero-pointing process. Our fiducial Fundamental Plane is presented in Section \ref{section:FP_results} alongside constancy checks for variations on our selection criteria and parameter estimates. Sections \ref{section:pv_cat} and \ref{section:conclusion} are a summary of our peculiar velocity catalogue and our conclusions, respectively.

Unless stated otherwise, we use assume a flat $\Lambda$CDM cosmological model with $\Omega_m = 0.3151$   \citep[Planck 2018 base plikHM+TTTEEE+lowl+lowE+lensing mean, including massive neutrinos;][]{Planck_Colab_2020_VI} and $H_0 = 100h\,\text{km}\,\text{s}^{-1}\text{Mpc}^{-1}$. All magnitudes use the AB system and all uses of `$\log$' refer to base 10 logarithms.

\section{Photometric and Spectroscopic Selection Criteria} \label{selection:primary}

The DESI Data Release 1 (DR1) Fundamental Plane (FP) peculiar velocity sample presented in this paper is extracted from a combination of photometric and spectroscopic data provided by the DESI Legacy Imaging Surveys DR9 \citep{Legacy_survey} and DESI DR1 data release \citep{DESI_collab_DR1}, respectively. The sample is primarily a subsample of the DESI Bright Galaxy Survey (BGS) sample, but also includes additional targets that were specifically targeted for the DESI peculiar velocity survey. Both surveys imposed baseline photometric selection criteria that are described in \citet{BGS_TS} and \citet{DESI_PV_target_selection}, respectively. 

In addition to these baseline target selection criteria, we impose an additional selection process to isolate a dispersion-supported early-type galaxy sample suitable for the Fundamental Plane analysis. The primary selection, described in the list below, is based on the criteria used in the SDSS peculiar velocity analysis \citep{Howlett_2022}. After measuring velocity dispersions (Section \ref{section:vdisp_measurments}) we apply further spectroscopic \citep{Said_EDR} to remove targets for which we were unable to obtain reliable velocity dispersion measurements. These additional cuts are described in detail in Section \ref{selection:vdisp_cuts}. An overview of all additional selection criteria, including the number of remaining targets after each cut, is presented in Table \ref{tab:selection_criteria}. 

Our primary selection criteria are as follows:
\begin{enumerate}[label=(\roman*)]
    \item objects are spectroscopically classified as galaxies (\texttt{SPECTYPE = "GALAXY"}); \\ \label{selection:galaxy}
    \item objects have a successful and robust redshift measurement with warning flag \texttt{ZWARN = 0} and \texttt{DELTACHI2 >30};\label{selection:quality_spec}
\end{enumerate}

\begin{table}[t]%[hbt!]
\begin{threeparttable}
 \caption{Selection criteria applied to the DESI DR1 data to create the DESI DR1 Fundamental Plane catalogue. In each row we provide a summary of each criteria, a reference to the location where it is described in the text, and a count of the number of remaining targets remaining in the sample after each criterion is applied.}
 \label{tab:selection_criteria}
 
 \begin{tabular}{lll}
 \toprule
 \headrow Selection Criteria & Reference& \# Remaining\\ 
 \midrule

 \texttt{SPECTYPE = "GALAXY"}& Section \ref{selection:primary} \ref{selection:galaxy} & 16,406,896\\ 
 \midrule 
 \texttt{ZWARN = 0} and \texttt{DELTACHI2 > 30}& Section \ref{selection:primary} \ref{selection:quality_spec} &13,977,157\\
 \midrule 
 \texttt{FLUX > 0} and \texttt{NOBS > 0}& Section \ref{selection:primary} \ref{selection:quality_phot} & 13,923,084\\ 
 \midrule 
  $0.0033<z<0.1$& Section \ref{selection:primary} \ref{selection:redshift}& 860,632\\ 
 \midrule 
 Magnitude in range $10.0 <m_r < 18.0$& Section \ref{selection:primary} \ref{selection:magnitude} & 336,299 \\ 
 \midrule 
 $g-r$, $r-z$ colour cuts& Section \ref{selection:primary} \ref{selection:colour} & 170,592 \\ 
 \midrule 
 de Vaucouleurs profile or S\'{e}rsic & \multirow{2}{*}{Section \ref{selection:primary} \ref{selection:profile} }&\multirow{2}{*}{116,085} \\ profile with $n_s > 2.5$\\
 \midrule 
 Axial ratio $b/a >0.3$& Section \ref{selection:primary} \ref{selection:axial} &110,794\\ 
 \midrule 
 Velocity dispersion in range & \multirow{2}{*}{Section \ref{selection:vdisp_cuts}}& \multirow{2}{*}{108,810}\\ $50.0<\sigma_{vdisp}<420$ and error>0.0 \\
 \midrule 
 Single measurement per galaxy& Section \ref{selection:dupes} & 98,292\\ 
 \midrule 
 No FP outliers& Section \ref{section:FP_fit} & 96,758 \\
 \midrule
 BGS Large-scale structure footprint & Section \ref{section:clustering_cat} & 73,822\\
 \bottomrule
 \end{tabular}
\end{threeparttable}
\end{table}

Here \texttt{SPECTYPE} is the spectral type (\texttt{GALAXY}, \texttt{QSO}, or \texttt{STAR}) of the best-fit redshift template from DESI pipeline \citep{Bailey_Redrock}. Requiring \texttt{SPECTYPE = "GALAXY"} removes the majority of stars and quasars from our sample. \texttt{ZWARN = 0} is a bitmask warning flag generated by the DESI pipeline that indicates if there was a problem with the redshift fit. \texttt{DELTACHI2} is the difference between the chi-squared values for the best two template fits, requiring \texttt{DELTACHI2 > 30} removes targets from our sample that had uncertain redshift measurements. Together these two cuts act as a spectroscopic quality control cut, removing targets that have poorly resolved spectra and spectra from multiple overlapping galaxies. 

\begin{enumerate}[resume*]
 \item objects have well defined flux measurements \texttt{FLUX > 0} and a positive number of photometric observations \texttt{NOBS > 0} in each of the g, r and z bands\label{selection:quality_phot};
\end{enumerate}

The photometric quality control cuts remove targets that have missing or problematic photometric data which can cause issues later in the Fundamental Plane analysis.

\begin{enumerate}[resume*]
 \item heliocentric redshift in the range $0.0033<z<0.1$\label{selection:redshift};
\end{enumerate}
We apply a redshift cut to limit our sample to the range where we can most accurately and reliably measure peculiar velocities. This range is bound by a lower redshift cut at $z=0.0033$ corresponding to a total velocity of $cz\approx 1000\,\kms$. This is sufficiently high that it is larger than the expected peculiar velocity measurements (a few hundred $\kms$), allowing peculiar velocity approximations like the \citet{Carreres_2023_pv_est} estimator used in this work to be applied, while still low enough to retain nearby galaxies with known distances for the zero-point calibration. Due to the $\sim20$\% intrinsic scatter in the FP relation, we expect peculiar velocity errors on the scale $0.2cz$. At our upper redshift bound of $z=0.1$, this corresponds to $\sim6000 \kms$ errors for individual peculiar velocity measurements. This is much larger then the expected measurements, but due to volume scaling and the large number of galaxies in our sample, peculiar velocity measurements near this upper limit still contribute information via population measurements of cosmology and bulk flows. It is important to note that the scaling of uncertainty proportional to redshift also applies to systematic uncertainties. Thus, even small systematics can have substantial impacts on results when propagated to large scales.

\begin{enumerate}[resume*]
 \item galactic extinction corrected, $r$-band apparent magnitude in the range $10.0 < m_r^{morph} < 18.0 $\label{selection:magnitude};
\end{enumerate}
The selection effect parameters used later in the fitting of the Fundamental Plane (Section \ref{section:FP_fit}) assume a magnitude-limited sample with limits imposed by this cut. Our sample contains targets selected by multiple survey teams. In the previously imposed redshift range of $0.0033<z<0.1$, this sample is primarily made up of BGS and PV targets with magnitude cuts of $m_r<20.0$ and $m_r<18.0$, respectively \citep{BGS_TS,DESI_PV_target_selection}. To ensure a clean cut for later use in the selection effect calculations, we use the tighter of the two cuts, $m_r<18.0$. No bright cut was implemented during target selection, so the cut used in SDSS \citep{Howlett_2022}, $m_r>10.0$, was used. Below this value, our photometry becomes sparse and unreliable, making it unsuitable for this analysis. The effect of this choice on the Fundamental Plane fit and subsequent distance measurements is investigated in Section \ref{section:IntValidation}. 

\begin{enumerate}[resume*]
 \item non k-corrected $g-r$ and $r-z$ colours in the region:
 \begin{align}
 &\bullet \,\,(m_g - m_r)>0.68,\label{ccut_1}\\
 &\bullet \,\,(m_g - m_r) > 1.3(m_r - m_z) - 0.05,\label{ccut_2}\\
 &\bullet \,\,(m_g - m_r)<2.0(m_r - m_z) - 0.15\label{ccut_3};
 \end{align}
 \label{selection:colour}
\end{enumerate}
Our colour cuts were calibrated during target selection \citep{DESI_PV_target_selection} to remove green valley and blue cloud galaxies (Equation \refeq{ccut_1}), dusty disk galaxies (Equation \refeq{ccut_2}), and  mergers, peculiar objects and imaging artifacts (Equation \refeq{ccut_3}). As this sample was drawn from the entirety of DR1 instead of just those targeted as part of the peculiar velocity spare fibre program, we have re-implemented these cuts to maintain consistency across the sample.

\begin{enumerate}[resume*]
 \item the highest likelihood model for the surface brightness profile fit of the object is the de Vaucouleurs profile (\texttt{TYPE = "DEV"}) or a S\'{e}rsic profile (\texttt{TYPE = "SER"}) with a S\'{e}rsic index $n_s > 2.5$\label{selection:profile};
 \item axial ratio $b/a$ greater than $0.3$.\label{selection:axial}
\end{enumerate}
Elliptical galaxies typically have a surface brightness distribution that follows the de Vaucouleurs profile. However, observational effects such as seeing and image noise can effect the likelihood of this fit. As such, we have relaxed this requirement to include galaxies that better fit the more general S\'{e}rsic profile, which allows for an additional free parameter, the S\'{e}rsic index $n_s$, to vary from $n_s=4$ characteristic of the de Vaucouleurs profile. The suitability of the relaxed requirement was investigated during target selection \citep[Figure 2]{DESI_PV_target_selection} where $n_s>2.5$ was identified as a good cutoff when evaluating the best fit S\'{e}rsic index for morphologically classified galaxies from the Siena Galaxy Atlas \citep[SGA,][]{SGA}. We re-evaluated the S\'{e}rsic index distribution for the DR1 sample and found results consistent with the target selection paper. Additionally, we apply an axial ratio cut of $b/a>0.3$ to remove edge-on disk galaxies (red spirals and lenticular galaxies) that remain in our sample despite the light-profile and colour cuts.

\section{Comparison of DR9 DECaLS and DR9 BASS/MzLS magnitudes}\label{sec:mag_offset_test}
Before finalising our input photometry, we investigate the consistency of our imaging. In previous papers, a discrepancy between DECaLS and BASS/MzLS $r$-band magnitudes was observed. A variety of different methods were used to characterise and correct for the offset between the measurements of the two surveys, including a colour dependent magnitude shift \citep{Zarrouk_2022}, a fixed offset that unifies the average target density across the two samples \citep{BGS_TS}, and a fixed offset based off the observed the median of the magnitude difference $m_r^\mathrm{BASS/MzLS} - m_r^\mathrm{DECaLS}$ \citep{Said_EDR}.

In this work, we choose to use a fixed offset with no colour dependence following \citet{BGS_TS} and \citet{Said_EDR}. However, as our sample is not yet complete in the overlapping region, it does not make sense to choose the offset based on what will make the target density consistent between the two samples, as done in \citet{BGS_TS}. The method used by \citet{Said_EDR} uses an external sample to identify the $m_r^\mathrm{BASS/MzLS} - m_r^\mathrm{DECaLS}$ magnitude difference. In this work, we only complete an internal comparison. The Legacy catalogue \citep{Legacy_survey} uses BASS/MzLS photometry for galaxies above the galactic plane and with declination $>32.375^\circ$, and DECaLS photometry otherwise. Using this split, we label galaxies as `BASS/MzLS primary' and `DECaLS primary' accordingly for our analysis.

To avoid the issues of sample size, we use a population model to measure the magnitude offset by fitting truncated Gaussian relations to each of the `DECaLS primary' and `BASS/ MzLS primary' subsets of our Fundamental Plane sample. As we are no longer directly comparing measurements between the two surveys we expand our above `primary' designation to the entire Fundamental Plane sample, rather than just the overlap region, assigning all galaxies above the galactic plane and with declination $>32.375^\circ$ as `BASS/MzLS primary',  and all other galaxies `DECaLS primary'. We then add an additional lower magnitude cut at $m_r>13$ mag to our selection criteria as brighter observations are sparse and proved problematic to the Gaussian fits. By applying our Fundamental Plane selection criteria prior to fitting, we remove possible discrepancies in the subsample populations due to variations in how targets from different sub-programs (Low-z, BGS, MWS) were observed across the sky. 

\begin{figure}[hbt!]
 \centering
 \includegraphics[width=\linewidth]{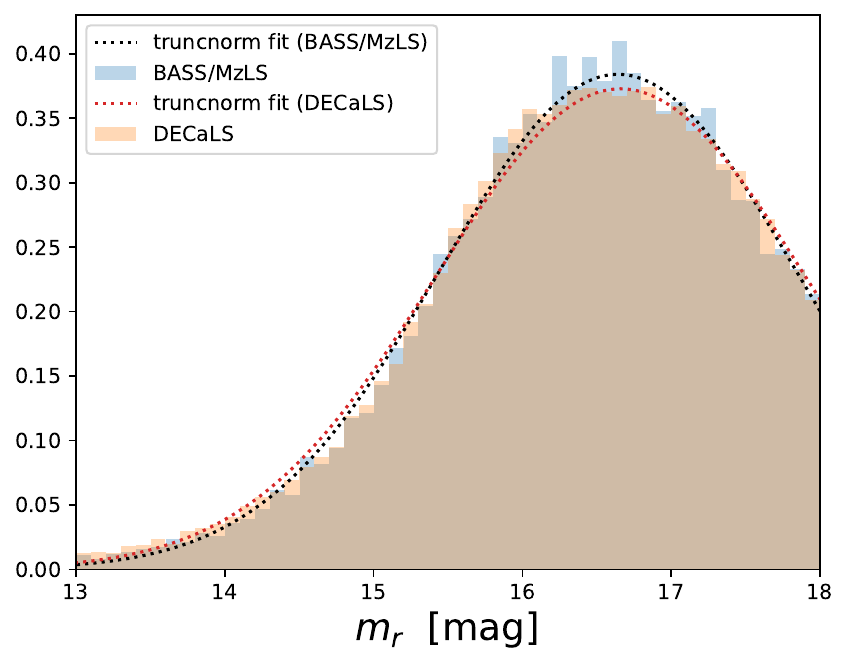}
 \caption{Apparent magnitudes and best-fit truncated Gaussian distribution for the DECaLS (orange) and BASS/MzLS (blue) `primary' subsets of the FP sample with magnitude cuts at $13<m_r<18$ mag. \red{The means of the two distributions are consistent to $1.4\sigma$, as such no magnitude correction is applied.}}
 
 \label{fig:BASS_DECaLS_offset}
\end{figure}

The best-fit Gaussians for each of the two survey subsamples are shown in Figure~\ref{fig:BASS_DECaLS_offset}. The best-fit mean and standard deviation for the `BASS/MzLS primary' and `DECaLS primary' subsamples are as follows:
\begin{align*}
    \mu_\mathrm{BASS/MzLS} &= 16.642^{+0.011}_{-0.012}, & 
    \sigma_\mathrm{BASS/MzLS} &= 1.191^{+0.012}_{-0.006},
    \\
    \mu_\mathrm{DECaLS} &=16.650^{+0.005}_{-0.009}, &
    \sigma_\mathrm{DECaLS} &= 1.249^{+0.005}_{-0.006}.
\end{align*}
This results in an offset $m_r^\mathrm{BASS/MzLS} - m_r^\mathrm{DECaLS} = -0.018 \pm 0.013$ which is consistent with zero at $1.4\sigma$. Thus, we do not apply any magnitude offset correction to our sample. 

The larger scatter in the DECaLS distribution, compared to BASS/MzLS, occurs primarily around Declination $\sim 0^\circ$. This region is observed with a higher pass count than the rest of the DR1 survey area. As such, incompleteness is believed to be the source of this discrepancy. We plan to reinvestigate this magnitude offset as part of the much more complete DR2 Fundamental Plane analysis to verify this assumption.

\section{Velocity Dispersions} \label{section:vdisp_measurments}

With our input photometry and spectra in hand, our next step is to obtain velocity dispersion measurements for our sample. To do so we employ Penalised Pixel-Fitting software \citep[pPXF, v8.2.4,][]{Cappellari_2004,Cappellari_2017,Cappellari_2023} to fit a combination of spectral templates to each DESI spectrum. For this analysis we used the Indo-US Coud\'e Feed Spectral library templates \citep{Valdes_2004} which have a spectral resolution of 1.35\AA, corresponding to a resolution limit of $\sigma=30\kms$ \citep{Valdes_2004}. There are two groups of DESI spectra: per-tile and full-depth \citep{DESI_collab_DR1}. In our analysis we used the full-depth spectra as these combine all available exposures for a given target. This increases the signal-to-noise ratio of the spectra and ultimately the likelihood of successfully measuring the velocity dispersion, which was identified during target selection to be a limiting factor of the Fundamental Plane peculiar velocity survey \citep{DESI_PV_target_selection}. 

\begin{figure}[hbt!]
 \centering
 \includegraphics[width=\linewidth]{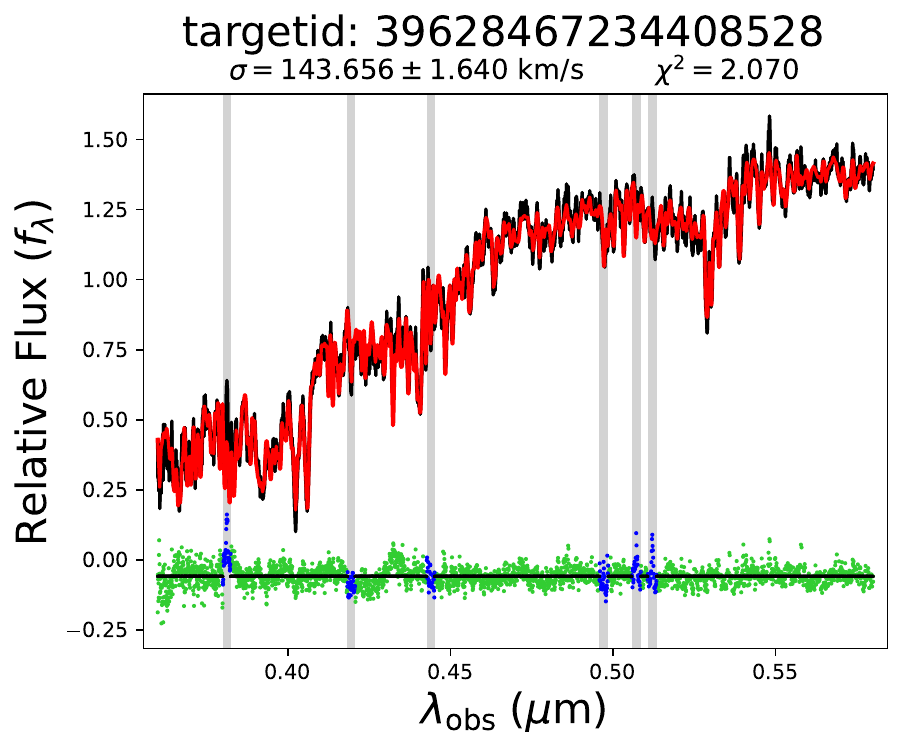}
 \caption{An example spectrum with velocity dispersion fit using pPXF. The plot shows the blue camera spectrum for the DESI \texttt{targetid 39628467234408528}. The spectrum is plotted as relative flux as a function of observation-frame wavelength, with `relative flux' referring to the observed flux normalised to the median flux of the spectrum. The observed spectrum is plotted in black with the best fit pPXF model plotted in red. In green and blue we show the residuals with the blue points indicating regions of the spectrum that were masked during the fit process as they contain known background emission wavelengths.}
 \label{fig:ppxf_spectra}
\end{figure}

Each DESI spectrum is split across three cameras: the blue-band spectrograph from $3600 - 5550$ \AA, the red-band spectrograph from $5550 - 6560$ \AA,  and the infrared spectrograph from $6560-9800$ \AA\  \citep{DESI_collab_2016b}. Historically, optical velocity dispersion measurements have mostly been made using a spectral range similar to that of the DESI blue-band camera. While this choice has primarily been due to instrumental limitations that are irrelevant to DESI, we have decided to continue use this range as it is the easiest to compare to the literature and to test for systematic effects. Additionally, in order to measure accurate velocity dispersions, pPXF requires that the resolution of the template library exceeds that of of the data. The Indo-US templates have a resolution limit of $\sigma=30\kms$ making them a good match for the blue camera with its limiting velocity dispersion measurement of $\sigma=46\kms$. However, as the red camera and z-band camera have resolution limits of $\sigma=31\kms$ and $\sigma=29\kms$ respectively, a new more resolved template library is required to get accurate velocity dispersions in these bands. Thus, we determined that measuring velocity dispersion from the blue-band spectra was the most suitable choice given current limitations. This may change for future analysis.  An example spectrum is shown in Figure~\ref{fig:ppxf_spectra}.

\subsection{Velocity dispersion cuts}
Once velocity dispersion measurements had been obtained, three additional cuts were applied to our sample. The first was a minimum velocity dispersion measurement of $\sigma=50\kms$\label{selection:vdisp_cuts}. This cut removes galaxies from our sample for which we were unable to successfully measure velocity dispersions. Additionally, galaxies with intrinsic velocity dispersions under or close to the resolution limit of the spectrograph can result in artificially inflated spectral velocity dispersions. Such measurements are unreliable, and so these galaxies were removed from our sample by setting the minimum required velocity dispersion slightly above the DESI blue camera resolution limit. The second cut we applied to our sample is a maximum velocity dispersion measurement of $\sigma = 420\kms$. This cut removes galaxies that had much higher velocity dispersion measurements than expected, usually the result of a poor or unsuccessful measurement. Finally, we require that the velocity dispersion error (from pPXF), $\delta\sigma >0$, as a zero-error measurement indicates that something failed during the fitting process.

\subsection{Primary observation identification}\label{selection:dupes}
As our sample consists of a compilation of all DR1 spectra that pass our selection cuts, many of the galaxies have been observed multiple times. The majority of these repeat observations are coadded by the DESI data reduction pipeline \citep{Guy_2023}. However, there are some repeat observations that are not merged, falling into 3 categories:
\begin{enumerate}
 \item the same \texttt{TARGETID} but different \texttt{PROGRAM} and/or \texttt{SURVEY},
 \item overlapping target location but different \texttt{TARGETID},
 \item and offset targets on the same galaxy. 
\end{enumerate}

Galaxies in the first category are usually observed in both dark and bright time. However, there are several galaxies that have been observed as `backup' targets (during poor observing conditions) before later being re-observed, as well as galaxies that have repeat observations across multiple survey stages --- sv1, sv2, sv3 and/or main, where sv are survey validation observations that were mostly released in the Early Data Release, EDR. In all situations, the spectra for each galaxy are coadded and saved separately for each \texttt{TARGETID}, \texttt{PROGRAM}, \texttt{SURVEY} combination. 

The second category occurs when a galaxy has been targeted separately by different surveys, usually BGS and peculiar velocity targets, with slightly different fibre positions. In this situation, it is possible that different surveys are trying to target different regions of the galaxy (e.g.\ the galaxy centre as well as a star forming region or orbiting dwarf galaxy). Due to their very small offsets of less than the fibre aperture radius of $\theta_{ap}=0.75$  \arcsec, there is a high percentage of overlap between repeat observations, resulting in both observations observing the same area on the sky. 

The third category usually occurs intentionally in large, nearby galaxies. In the peculiar velocity target selection, some of the FP galaxies were specifically targeted multiple times using different fibre offsets to assist in understanding systematic effects associated with non-central fibre placement \citep[`extended' FP targets,][]{DESI_PV_target_selection}. Additionally, other surveys have targeted different points of interest such as star-forming regions, supernovae, etc. within nearby galaxies, some of which may have made it into our sample. This third category of repeat observations are usually low priority and are only observed when there are no other targets available within the fibre patrol radius.

\begin{figure}[t!]
 \centering
 \includegraphics[width=\linewidth]{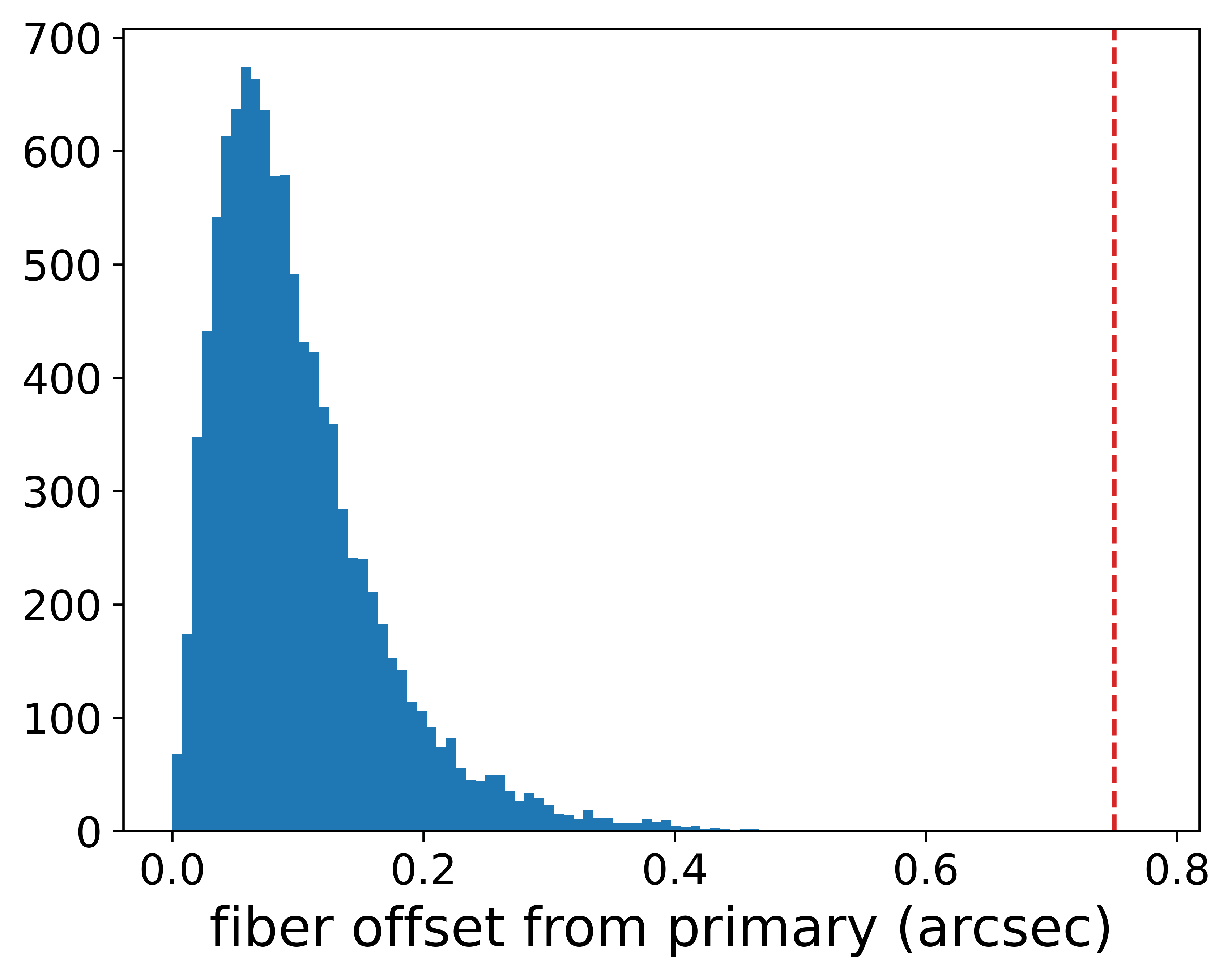}
 \caption{Distribution of on-sky separation (in  \arcsec) between duplicate and \texttt{primaryVdisp} fibre locations for repeat observations of the same galaxy. The red dashed line is the on-sky fibre radius \red{($\theta_{\text{ap}} = 0.75$ arcsec)} for DESI fibres. All but one duplicate observation overlaps with the primary observation. }
 \label{fig:fibre_offset}
\end{figure}

Figure~\ref{fig:fibre_offset} shows the distribution of the separations between primary and duplicate fibre locations. Almost all duplicate observations have significant overlap with the chosen `primary' observation for the galaxy, with separations much less than the fibre radius (red line). As such, the majority of repeat observations fall into either the first or second categories. This happens when a galaxy has been observed with slightly different fibre positions across different \texttt{SURVEY/PROGRAM/TARGETID} combinations, either due to repeatability of the fibre positioning or slightly different targeting coordinates. In our sample, only one of the 10,492 repeatedly observed galaxies falls into category 3; the primary and secondary observations for this galaxy are offset by $0.78$ arcsec which corresponds to an overlap of $40\%$ of the primary fibre-area. The reason so few category 3 repeats exist in the DR1 sample is likely due to the low-priority of extended PV targets, which are yet to be observed, and the light-profile and colour cuts that remove star-forming, clumpy and disrupted galaxies targeted by other projects. As such, contamination from significantly offset observations is not an issue for the DR1 FP catalogue but will become increasingly important to identify and analyse such galaxies in future data releases as the survey nears completion.

As all but one of our repeat observations are have very high overlap, we treat them as if they are the measuring the same region of the galaxy. We chose the observation with the best relative velocity dispersion error as the primary observation, labelling them as such using the \texttt{primaryVdisp} flag. We use the repeated observations to perform internal consistency checks on our velocity dispersion measurements (Section \ref{section:vdisp_internal_consistancy}), but only the primary observations are used when doing external consistency checks (Section \ref{section:vdisp_external_consistancy}), when fitting the Fundamental Plane parameters (Section \ref{section:FP_fit}), and during zero pointing (Section \ref{section:zp_and_pvs}). However, all observations are included in the final catalogue with independently calculated log-distance and peculiar velocity measurements to allow for easy crossmatching to other catalogues.\\

\subsection{Internal consistency}\label{section:vdisp_internal_consistancy}
As the Fundamental Plane calibration is performed unanchored, it is not crucial that any of the 3 FP parameters is consistent with other surveys, as any offsets will be corrected during zero-pointing. More complex deviations, such as tilts, bends or redshift-dependent and coordinate-dependent effects are not removed during this process and need to be identified. As such, consistency of measurements across subsamples is essential when they represent the same population of galaxies. The most obvious way to select such a subsample is by observing conditions (bright vs dark time) and by position on the sky. Additionally, we check that we are accurately estimating the velocity dispersion errors by comparing primary and secondary observations as described in Section \ref{selection:dupes}.

\begin{figure}[t!]
 \centering
 \includegraphics[width=\linewidth]{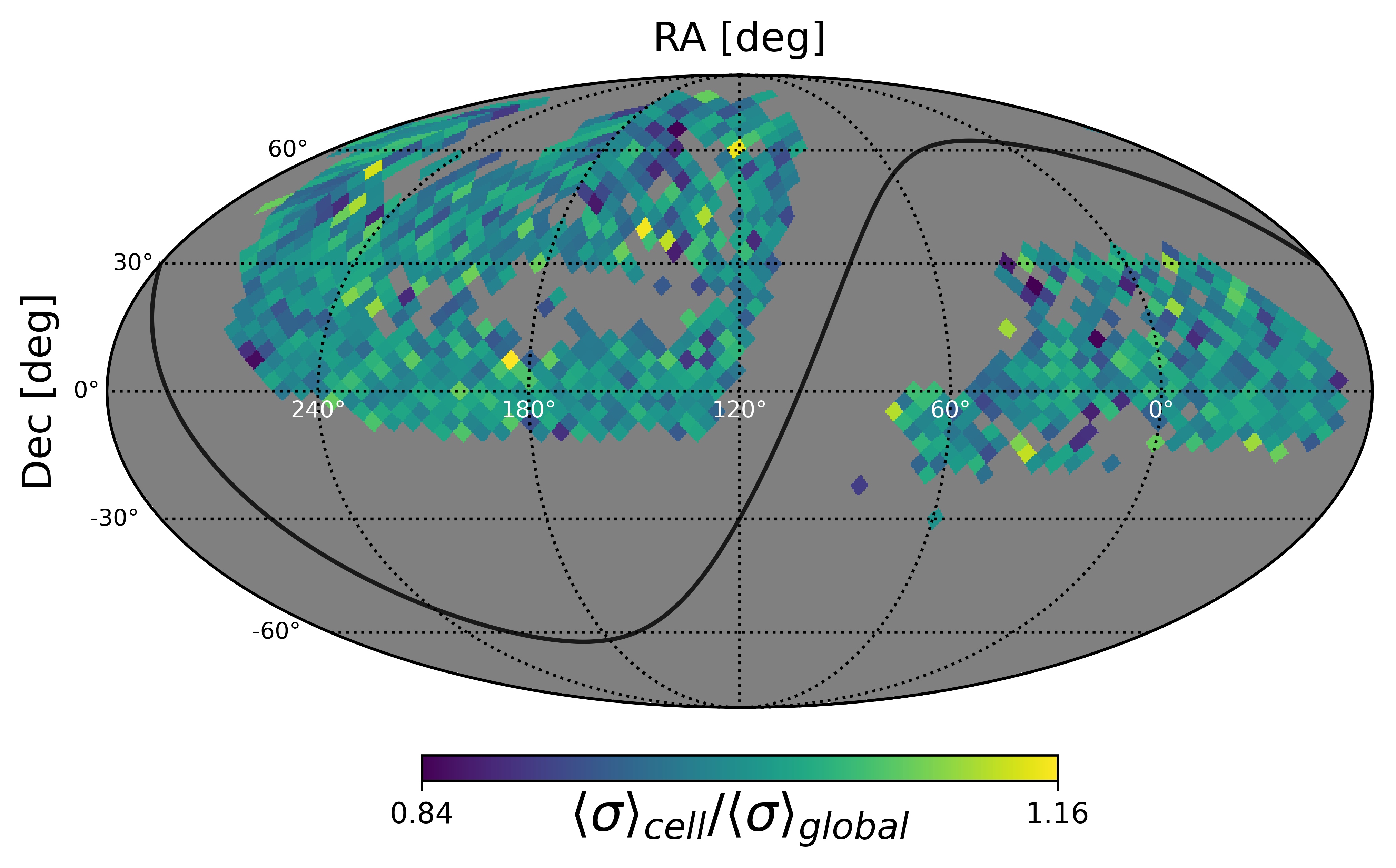}
 \caption{Variation of velocity dispersion measurements across the sky for cells with more than 20 observed galaxies. Each cell is colour coded based of the mean velocity dispersion of the cell which is then normalised to the global mean velocity dispersion. \red{The cell standard deviation is $5\%$ across the $1156$ cells with observations. With the maximum cell offset at $16\%$ and no correlations with sky location there is no evidence of bias.}}
 \label{fig:skymap_vdisp_consitancy}
\end{figure}

In Figure \ref{fig:skymap_vdisp_consitancy}, we check for deviations in velocity dispersion measurements across a variety of different sky positions. We do this in HEALPix bins of width $\sim3.7^\circ$ \citep{Gorski_2005}. Masking out bins with 20 observations, we calculate the mean velocity dispersion in each bin and colour-code by the percent offset from the global mean. In the DR1 data, are no noticeable patterns in the figure based off sky-position.

\begin{figure}[hbt!]
 \centering
 \includegraphics[width=\linewidth]{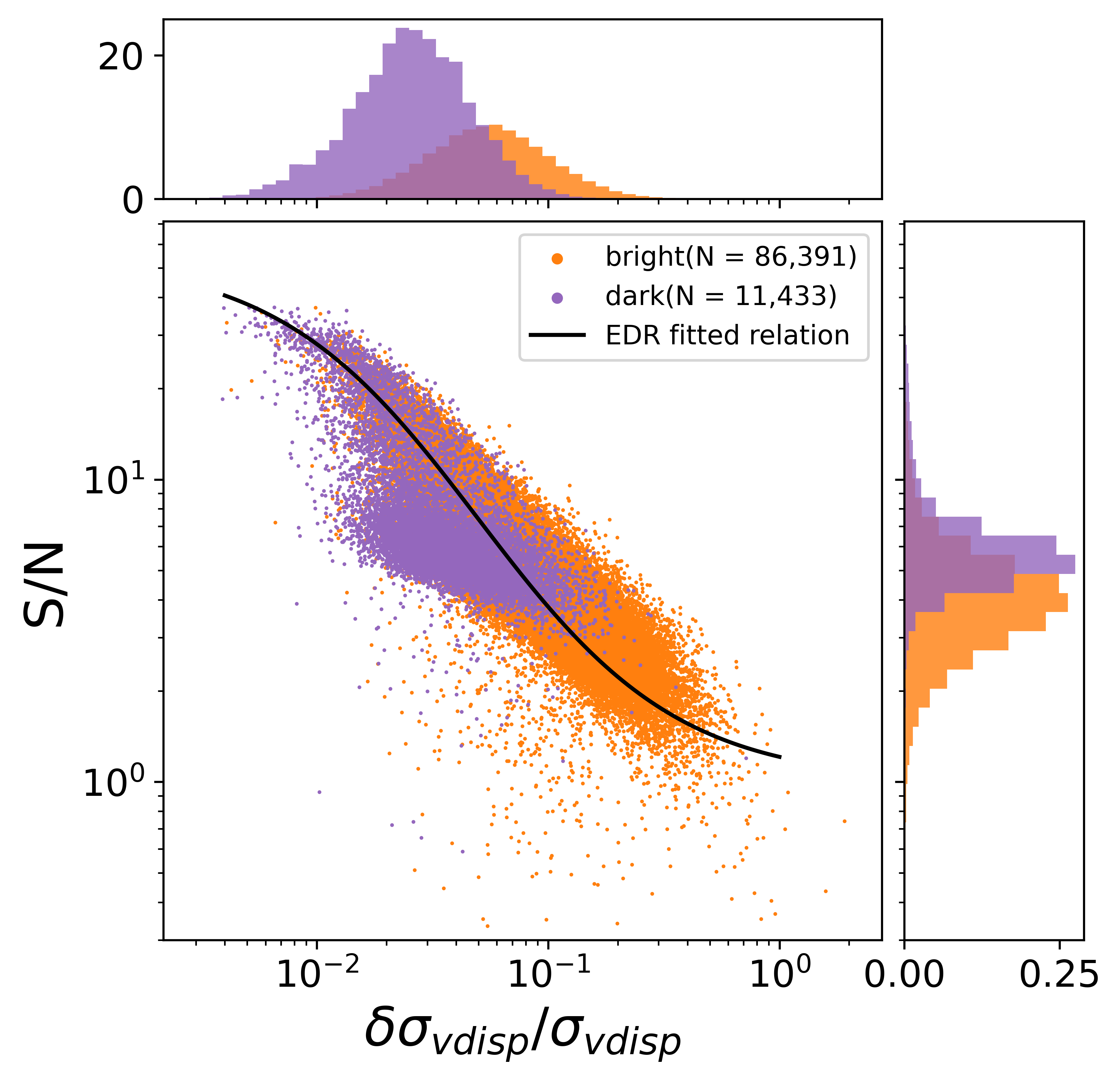}
 \caption{Relative velocity dispersion error vs SNR for both bright (orange) and dark time (purple) observations. The bright time observations are consistent with the early data release (EDR) relation observed in \citet[over-plotted in black]{Said_EDR} with the dark time observations on average obtaining tighter constraints on the velocity dispersions measurements compared to equivalent SNR bright time observations.}
 \label{fig:bd_snr_vdisp}
\end{figure}

Next, we analyse dependencies between bright and dark time observations. In Figure \ref{fig:bd_snr_vdisp} we look at the relationship between the signal to noise ratio (SNR) and the relative velocity dispersion errors for both subsamples. From the plot we can see that the bright-time observations (orange) are consistent with the relation (black line) found by \citet{Said_EDR}. When looking at the dark-time observations (purple) the measurements have higher SNR and smaller relative velocity dispersion errors when compared to bright observations with a similar SNR.

\begin{figure*}[t]
 \centering
 \includegraphics[width=\linewidth]{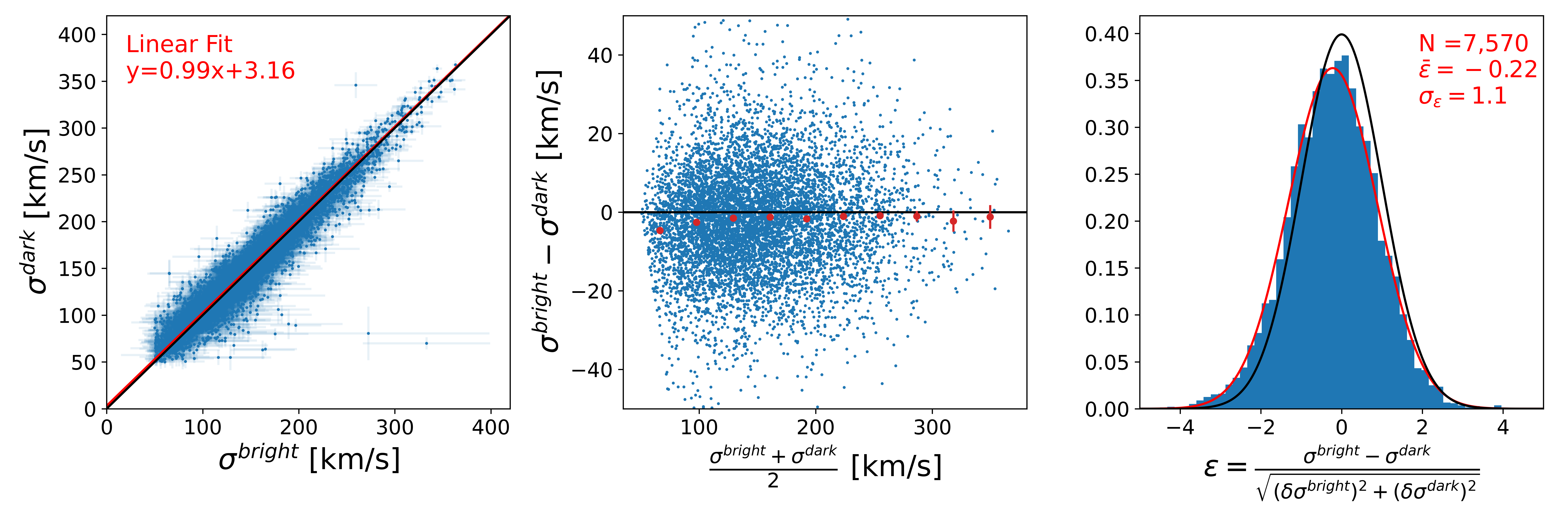}
 \caption{Pairwise, galaxy by galaxy comparison between velocity dispersion measurements obtained from bright and dark time spectra. Left panel: One-to-one comparison of pPXF velocity dispersion measurements with best-fit linear relation (red) and one-to-one line (black). Middle panel: Difference between the measurements as a function of the mean velocity dispersion. Over-plotted in red are the sample means in bins. Right panel: Distribution of the velocity dispersion measurement pairwise relative errors, with best-fit Gaussian (red) and unit Gaussian distribution (black). There is an offset of $\bar{\varepsilon} = -0.22$ indicating that dark time velocity dispersion measurements tend to be larger then bright time after accounting for errors. To correct for this we apply a $+2.3\pm0.1\kms$ correction to all bright-time velocity dispersion measurements.}
 \label{fig:vdisp_BrightDark_comp}
\end{figure*}

In Figure \ref{fig:vdisp_BrightDark_comp}, we directly compare the velocity dispersion measurements made in bright and dark time using the $7570$ galaxies with observations in both programs. We find that there is an offset between the bright and dark samples with the dark samples measuring larger velocity dispersions. When weighting by the combined errors, this offset is $\bar{\varepsilon} = -0.22$ (i.e. $22\%$ of the contoured error). We correct for this offset by applying a $+2.3\pm0.1\kms$ correction to all bright-time velocity dispersion measurements, calculated by taking the variance-weighted mean of the offset, which brings them into better consistency with the dark-time measurements. The source of this offset will be further investigated in the DR2 analysis.

\begin{figure*}[t]
 \centering
 \includegraphics[width=\linewidth]{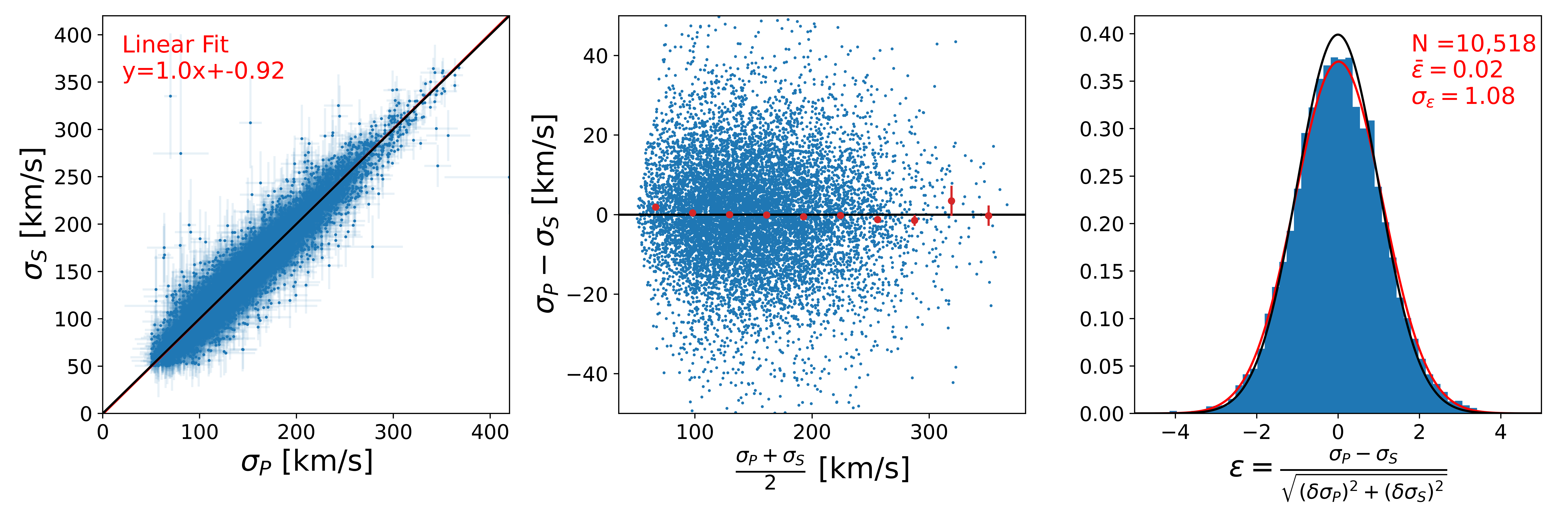}
 \caption{As in Figure \ref{fig:vdisp_BrightDark_comp} but between velocity dispersion measurements obtained from `primary' and `secondary' spectra after applying the bright-dark correction. This sample consists of all repeated observations in our Fundamental Plane sample. The best fit Gaussian distribution (red line, right panel) is given by $\bar{\varepsilon}= 0.02$ $\sigma_\varepsilon =1.08$, indicating good agreement between primary and secondary measurement but an underestimation of the velocity dispersion errors which we account for by scaling the velocity dispersion errors of our full sample by 1.13.}
 \label{fig:vdisp_p_v_s_comp}
\end{figure*}

After applying this correction, we compare the the primary and secondary velocity dispersion measurements to check the velocity dispersion error calculations. Figure \ref{fig:vdisp_p_v_s_comp} shows that the primary and secondary velocity dispersion measurements are consistent ($\bar{\varepsilon}= 0.02$) but the errors are underestimated slightly ($\sigma_\varepsilon =1.08$). The slightly higher primary measurements indicated by $\bar{\varepsilon}>0$ are due to our \texttt{primaryVdisp} selection criterion being the smallest relative error; if two observations have the same measurement error, we take the largest measurement. Since this offset is small, it should not cause any significant biases for the DR1 sample, but we will revisit using an alternative criteria for \texttt{primaryVdisp} in the DR2 analysis. Finally, we correct for our underestimated errors by manually inflating our velocity dispersion errors across our entire sample by a factor of 1.13 so that $\sigma_\varepsilon =1.0$.

\begin{figure*}[t]
 \centering
 \includegraphics[width=\textwidth]{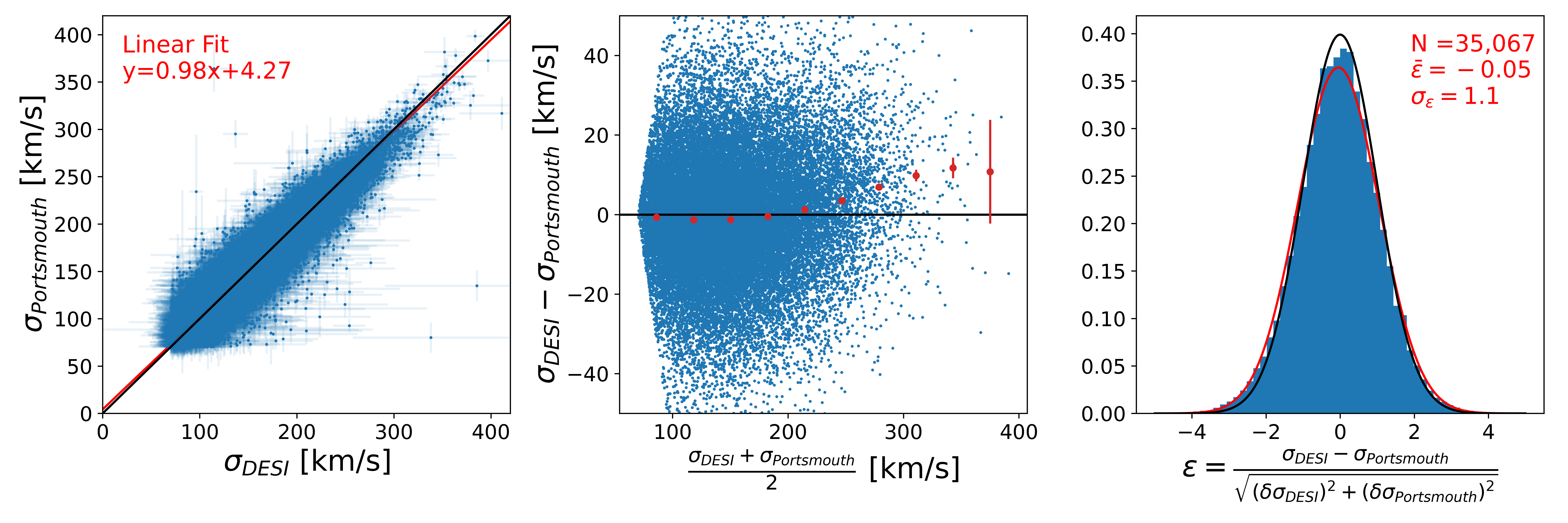}
 \caption{As in Figure \ref{fig:vdisp_p_v_s_comp} but between DESI measurements (this work) and those from the Portsmouth catalogue \citep[SDSS,][]{Portsmouth_cat}. All DESI measurements are post corrections (as discussed in Section \ref{section:vdisp_internal_consistancy}) and in the case of multiple observations, we use the `primary' observation. \red{Overall, there is good agreement between the two catalogues with the overall offset within 5\% of errors. A small discrepancy ($\lesssim 5\%$) emerges at high velocity dispersion ($\sigma>250\kms$) as data becomes sparse, but this has minimal impact on the overall sample agreement.} }
 \label{fig:DESI_SDSS_Vdisp_ppxf}
\end{figure*}

\subsection{External consistency} \label{section:vdisp_external_consistancy}
To validate the consistency of our velocity dispersion measurements, we compared velocity dispersions measured using DESI spectra to those measured in SDSS DR16 for galaxies present in both samples. This was done by crossmatching our entire Fundamental Plane sample to SDSS DR16 using the external match files (\texttt{survey-dr9-<region>-specObj-dr16.fits}) provided by the Legacy Surveys as well as the north/south region of each galaxy and the \texttt{BRICKID} and \texttt{BRICK\_OBJID} parameters. From this crossmatch we obtain SDSS object IDs as well as pipeline velocity dispersion measurements. We then further crossmatched using the internal SDSS IDs to the Portsmouth group catalogue \citep{Portsmouth_cat} which provides pPXF velocity dispersion measurements. This process leaves us with a crossmatched sample of $37,038$ unique galaxies with both DESI (this work) and SDSS (Portsmouth catalogue) velocity dispersion measurements. Before completing any analysis, we apply the SDSS velocity dispersion cuts $70 \kms <\sigma_{vdisp}<420 \kms$ \citep{Howlett_2022} on both the DESI and SDSS measurements, further restricting our sample to $35,067$ galaxies with valid velocity dispersion measurements.

The Portsmouth velocity dispersion measurements were obtained using the pPXF method, the MILES stellar library \citep{Sanchez_2006} and SDSS spectra. The method is the same as used in this work. The pPXF fitting code has been extensively tested using simulations \citep{Cappellari_2004, Cappellari_2017} and compared against measurements obtained using other methods in the SAURON \citep{Cappellari_2004} and LEGA-C \citep{Cappellari_2023} surveys where the method has been shown to consistently reproduce galaxy velocity dispersions. As the DESI and SDSS instrumentation and analysis pipelines are otherwise independent, we treat the respective velocity dispersions as being fully independent observations.

In Figure \ref{fig:DESI_SDSS_Vdisp_ppxf}, we compare the velocity dispersion measurements obtained by this work and \citet{Portsmouth_cat}. Both sets of measurements are in good agreement, with a preference for slightly smaller velocity dispersions measured by DESI but within $5\%$ of the contoured errors. When analysing the differences in measurement bins (middle panel), the DESI and \citeauthor{Portsmouth_cat} measurements are very consistent until $\sigma \approx 250 \kms$, at which point observations begin to become more sparse and the DESI measurements tend to be slightly larger than those of \citeauthor{Portsmouth_cat} This is a shift in the opposite direction to what is seen in the sample as a whole. The effect of these high velocity dispersion measurements on the mean offset is minimal, with the disagreement only increasing to within $7\%$ of the contoured error bars when removing all galaxies with either velocity dispersion measurement greater than $250\kms$. Both of these comparisons are much more consistent then what was observed during EDR \citep{Said_EDR} where consistency with SDSS was only within $18\%$ of the contoured errors. 

When looking at the error measurements in Figure \ref{fig:DESI_SDSS_Vdisp_ppxf} (right panel) we obtain $\sigma_{\epsilon} = 1.1$ indicating a slight underestimate in one of or both of the DESI and SDSS errors. Unfortunately, with our current analysis it is impossible to determine which of the two samples (or both) are underestimating errors. It is important to note that, although this issue is not completely resolved, the combined error underestimate has been reduced significantly from the $\sigma_{\epsilon} = 1.17$ that was measured by \citet{Said_EDR} in DESI EDR.

\section{Fitting the Fundamental Plane}
Following our initial selection criteria presented in the previous two sections, we fit a Fundamental Plane relation using a method modified from that presented in \cite{Howlett_2022}. From the Legacy Survey photometric morphological profile fits we obtain $r$-band half light radius $r^\text{morph}$, $g$, $r$ and $z$ band galactic extinction corrected flux $f_g$, $f_r$ and $f_z$, and perpendicular ellipticity components $\epsilon_1$ and $\epsilon_2$. From these we can compute ellipticity
\begin{equation}
 \epsilon = \epsilon_1 +i \epsilon_2,
\end{equation}
as well as the axial ratio
\begin{equation}
 b/a = \frac{1-|\epsilon|}{1+|\epsilon|}.
\end{equation}
From spectroscopy, we obtain heliocentric redshifts $z$ from the DESI data reduction pipeline, from which we calculate CMB-frame redshifts $z_{\rm cmb}$ assuming the CMB-dipole as measured by \cite{Planck_Colab_2020_I} and following the method from \cite{Carr_2021,Carr_2022}. We also take the velocity dispersions as described in Section \ref{section:vdisp_measurments}.

From these parameters we are able to calibrate the Fundamental Plane relation to our sample. This relation is of the form
\begin{equation}
 \log R_e = a\log\sigma_0 + b\log I_e + c \label{FP_trad}
\end{equation}
where $R_e$ is the effective radius ($\hikpc$), obtained from the angular effective radius $\theta_e$ and the distance; $\sigma_0$ and $I_e$ are the distance independent parameters of centralised velocity dispersion ($\kms$) and effective surface brightness ($\rm{L}_\odot \rm{pc}^{-2}$) respectively; and $a$, $b$, and $c$ are calibration parameters for the Fundamental Plane relation. For the remainder of this paper we adopt the common shorthand notation $r_z = as +bi + c$, where $r_z = \log R_e$, $s=\log {\sigma_0}$ and $i = \log I_e$.

\subsection{Parameters}\label{section:FP_parameters}

We compute an angular effective radius from the half-light radius $r^\text{morph}$ and axial ratio $b/a$ using the formula:
\begin{equation}
 \theta_e = r^\text{morph}\sqrt{b/a}. 
\end{equation}
From this we can calculate an observed physical effective radius, $r_{z}$, by using the observed CMB-frame redshifts as a proxy to calculate comoving distances, $d(z_{cmb})$, assuming no peculiar velocities. This gives us the first distance-dependent Fundamental Plane parameter in units of $\hikpc$: 
\begin{multline}
 r_z = \log(\theta_e) + \log(d(z_\text{cmb})) - \log(1+z_\text{helio}) \\+\log\left(\frac{1000\pi}{180 \times 3600}\right).\label{eq:FP_r}
\end{multline}
We note that this observed physical radius, $r_{z}$, will differ from the true effective radius of the galaxy due to the presence of peculiar velocities, which is what later enables us to extract the peculiar velocity of each galaxy given a calibrated FP relation. More concretely, given the fitted Fundamental Plane, we can define
\begin{equation}
 r_{z} - as - bi - c = \log(d(z_\text{cmb})) - log(d(\bar{z})) \equiv \eta,
\end{equation}
where $d(\bar{z})$ is the true comoving distance to the galaxy (with cosmological redshift $\bar{z}$) and $\eta$ is the so-called log-distance ratio.

The second Fundamental Plane parameter, $s$, is calculated from the velocity dispersions described in Section \ref{section:vdisp_measurments}. However, these measurements are obtained by running pPXF on spectra measured through a fixed diametre fibre. To be used as a distance-independent parameter in the Fundamental Plane, we need to convert this measurement into the intrinsic central velocity dispersion for each galaxy by centralising the observed measurement. This is done because comparably sized galaxies at the same distance will have similar intrinsic velocity dispersions. However, when one of these galaxies is further away, it will appear smaller in the sky, and thus a larger proportion of the galaxy will be covered by the fibre, resulting in a decreased velocity dispersion measurement. Similarly, different sized galaxies at the same distance will have different proportions of their light captured by the fibre, artificially decreasing the measured velocity dispersion of the smaller galaxy. By centralising the velocity dispersion measurements we remove the effect of the angular size of each galaxy, making our measurements consistent across our sample.

To obtain the centralised velocity dispersion measurements, we apply the following aperture correction:
\begin{equation}
 s = \log(\sigma) + \alpha_{ap}\big[\log(\theta_e) - \log(8\theta_{ap})\big], \label{eq:FP_s}
\end{equation}
where $s$ is the centralised velocity dispersion in $\kms$, $\sigma$ is the measured velocity dispersion obtained from the pPXF fit described in Section \ref{section:vdisp_measurments}, $\theta_e$ is the angular effective radius, $\theta_{ap} = 0.75$ is the aperture radius (in \arcsec) of the DESI fibres \citep{DESI_collab_2022_Inst}  and $\alpha_{ap}$ is an aperture correction parameter. 

A variety of aperture correction parameters $\alpha_{ap}$ have been reported in the literature, ranging from $\alpha_{ap} = -0.066\pm 0.035$ \citep{Cappellari_2006} to $\alpha_{ap} = -0.033\pm 0.003$ \citep{deGraaff_2021} with varying sample selection criteria. The choice of this parameter makes little difference to the individual log-distance measurements, causing shifts on the order of $\Delta \eta \approx 0.005$, well within the typical Fundamental Plane-derived error of $\sigma_\eta \approx 0.12$ dex. However, as is discussed in Section \ref{section:internal_effect_alpha}, across the entire sample the choice of aperture correction can introduce a redshift-dependent tilt. For our baseline analysis, we used $\alpha_{ap} = -0.06 \pm 0.03$ which flattens this tilt while encapsulating the range of literature values measured from early-type galaxy samples.

The last distance-independent Fundamental Plane parameter is the effective surface brightness. This parameter is measured from the apparent magnitude and angular effective radius of each galaxy
\begin{align}
 i = 0.4(M^r_\odot - m_r^\text{morph} - Qz_\text{cmb} + k_r)- \log(2\pi\theta_e^2) &\notag\\ 
 + 4\log(1+z_\text{helio}) + 2\log(64800/\pi )& \label{eq:FP_i}
\end{align}
where $M^r_\odot= 4.65$mag is the $r$-band absolute solar magnitude, $m_r^\text{morph} = \log(f_r/A_r)$ is the galactic-extinction-corrected, $r$-band apparent magnitude for each galaxy, $k_r$ is the $r$-band k-correction\footnote{We used \texttt{KCORR01\_SDSS\_R} from \citet{fastspecfit}.} and $2\log(64800/\pi)$ is a unit conversion term. We use the evolution correction parameter of $Q=1.1\pm0.4$, which is calculated by taking the mean of the values found in the literature \citep{Bernardi_2003, Blanton_2003, Loveday_2012, Loveday_2014, McNaught_2014,  Loveday_2015} with error equal to the standard deviation of these values. The effects of varying this correction are explored Section \ref{section:evo_correction}.

\subsection{Fitting Methodology}\label{section:fit_methodology}

To obtain log-distance ratios and peculiar velocities for our Fundamental Plane sample, we fit a 3D Gaussian Fundamental Plane model using a maximum likelihood method. This method was initially formulated by \citet{Saglia_2001} and \citet{Colless_2001} for the EFAR Fundamental Plane analysis but has been further refined in subsequent Fundamental Plane studies \citep{Magoulas_2012, Springob_2014, Said_2020, Howlett_2022, Said_EDR}. The method is a three-step process that first fits a Fundamental Plane model assuming no peculiar velocities, then uses the obtained Fundamental Plane parameters to fit the most likely log-distance ratio to each galaxy individually. After the fitting process is complete, the sample is anchored to a subsample of galaxies with known distances, returning any (if present) radial bulk-flow motion to the sample. \footnote{The initial fitting process is relative (between galaxies) and assumes that over a large volume no radial bulk flow is present}

The likelihood of our Fundamental Plane model for $N$ galaxies is:
\begin{multline}
 \mathcal{L} = \prod^N_{n=1} \Bigg( \frac{1}{(2\pi)^{3/2} \lvert\textbf{C}_n\rvert^{1/2}f_n} \\\times\exp\left[-\frac{1}{2}(\bm{x}_n -\bm{\bar{x}})\textbf{C}_n^{-1}(\bm{x}_n - \bm{\bar{x}})^T\right] \Bigg)^{1/S_n} \label{eq:FP_likelihood},
\end{multline}
where $\bm{x}_n = \{r_n,s_n,i_n\}$ are the Fundamental Plane parameters (as described in Section \ref{section:FP_parameters}) for each observed galaxy $n$, $S_n$ is an inverse weighting based off the commonly used $1/V_{\text{max}}$ weighting \citep{Schmidt_1968} and accounts for galaxies missing from our sample due to the redshift, magnitude and velocity dispersion cuts through an up-weighting of the galaxies within the sample. Here $\bm{\bar{x}} = \{\bar{r}, \bar{s}, \bar{i}\}$ are the mean values of the Fundamental Plane model and $\mathbf{C}_n$ is the covariance matrix describing both the intrinsic scatter and measurement uncertainty in the Fundamental Plane. This can be determined from the measured $r$, $s$ and $i$ uncertainties, two Fundamental Plane coefficients $a$ and $b$, as well as three orthogonal intrinsic scatter parameters $\sigma_1$, $\sigma_2$ and $\sigma_3$. The $f_n$ function is a normalisation term that corrects for the `selection bias' of each individual galaxy and has been modified from \citet{Howlett_2022} to include an upper velocity dispersion cut. It can be found below by substituting in the minimum and maximum velocity dispersion and magnitude cuts in log-space (i.e. for this work $s_\text{min} = \log(50)$, $s_\text{max} = \log(420)$, $m_\text{min} = 10$, and $m_\text{max} = 18$):
\begin{multline}
 f_n = \frac{1}{(2\pi)^{3/2}\lvert\bm{C}_n\rvert^{1/2}}\int_{-\infty}^{\infty}\int_{r_\text{min}-i/2}^{r_\text{max}-i/2}\int_{s_\text{min}}^{s_\text{max}}\text{d}s\,\text{d}r\,\text{d}t \\
 \times \exp\left[-\frac{1}{2}(\bm{x_n}-\bm{\bar{x}})\bm{C}_n^{-1}(\bm{x_n}-\bm{\bar{x}})^T\right]\label{eq:updated_fn_calc}
\end{multline}
where,
\begin{multline}
    r_\text{min(max)} = \frac{1}{5}\big[10 + M_\odot^r + 5\log(1+z_\text{helio}) - Qz_\text{cmb}\\-2.5\log(2\pi) + k_r + 5\log(d(\bar{z}))-m_\text{min(max)}\big] \label{eq:r_min/max}
\end{multline}
During this calculation, $f_n$ is calculated individually for each galaxy at each proposed cosmological redshift $\bar{z}$ and comoving distance $d(\bar{z})$. As a result, each galaxy has many $f_n$ values that vary as the galaxy's proposed distance increases. A full derivation of equations \ref{eq:FP_likelihood}, \ref{eq:updated_fn_calc} and \ref{eq:r_min/max} along with contained terms, can be found in \citet{Howlett_2022}. The aforementioned paper describes a method to reduce the $f_n$ integral to a sum of elementary functions for fast computation. Using these equations, we are left with an 8-parameter model ($a,\, b,\, \bar{r},\, \bar{s},\, \bar{i},\, \sigma_1,\, \sigma_2,\, \sigma_3$) which we fit to our data to obtain the Fundamental Plane relation for our dataset.

\subsubsection{Fitting the Fundamental Plane}\label{section:FP_fit}
To fit the Fundamental Plane model we first remove all repeat observations from our sample by limiting our sample to \texttt{primaryVdisp = True}, reducing our sample to $98,292$ unique galaxies. We then use the likelihood function in equation (\refeq{eq:FP_likelihood}) to fit the Fundamental Plane. The maximisation of this likelihood is done using the SciPy differential evolution optimisation algorithm \citep{Storn_Price_1997,SciPy_Virtanen_2020} with a tolerance of $10^{-6}$ and a max iteration count of $10,000$. Uniform priors were set on each of the $8$ free parameters with limits as shown in Table \ref{tab:paramater_priors}.

\begin{table}[ht!]
    \centering
    \small
    \begin{tabular}{lc}
       \toprule
       Parameter & Prior \\\hline
       $a$ & $\mathcal{U}(1.0, 1.8)$ \\
       $b$ & $\mathcal{U}(-1.5, -0.5)$ \\
       $\bar{r}$ & $\mathcal{U}(-0.5, 0.5)$ \\
       $\bar{s}$ & $\mathcal{U}(2.0, 2.4)$ \\
       $\bar{i}$ & $\mathcal{U}(2.4, 3.0)$ \\
       $\sigma_1$ & $\mathcal{U}(0.01, 0.12)$ \\
       $\sigma_2$ & $\mathcal{U}(0.05, 0.5)$ \\
       $\sigma_3$ & $\mathcal{U}(0.1, 0.3)$ \\\bottomrule
    \end{tabular}
    \caption{Uniform priors for each of the 8 fitted parameters of the Fundamental Plane model, where $\mathcal{U}(a,b)$ denotes a uniform distribution between $a$ and $b$.}
    \label{tab:paramater_priors}
\end{table}

Following the fit, we compute the $\chi^2$ difference between each data point and the best fitting Fundamental Plane model, from which we calculate the $p$-value for whether or not each galaxy is part of the distribution. Galaxies that have $p<0.01$ are identified as outliers and removed from our sample. The entire fitting process is then repeated until the total number of outliers remains constant, taking the final iteration as our fitted model. This procedure converged after 6 iterations, removing $1,534$ outlier galaxies from our sample, leaving us with $96,758$ calibration galaxies which are flagged as \texttt{FPcalibrator} in our final PV catalogue.

\subsubsection{Fitting the Log-Distance Ratios} \label{section:logdist_fit}
As described in Section \ref{section:FP_parameters} the $r_z$ values used as inputs to the Fundamental Plane model represent the distance-dependent parameter of the Fundamental Plane. When computing these values we use redshift as a proxy for distance. However, the true distance is required in order to determine galaxy peculiar velocities. Using the Fundamental Plane parameters calculated in the previous step, we can estimate the true intrinsic size of the galaxy $r_t$. The difference between the measured and estimated sizes is equal to the log-distance ratio;
\begin{equation}
    r_z-r_t = \eta \equiv \log(d(z_{cmb})/d(z_t)).
\end{equation} 
This estimate of the true distance of the galaxy allows the peculiar velocity to be determined.

We fit the log-distance ratios by generating an array of 1001 uniformly distributed candidate log-distance ratios in the range $[-1.5,1.5]$ for each galaxy. We compute the likelihood for each log-distance and target combination using equation (\ref{eq:FP_likelihood}), fixing $N = 1$, $S_n = 1$ and $\bm{\bar{x}}$ and $\textbf{C}_n$ to the best-fitting parameters from the previous step. This likelihood is combined with a flat prior on the log-distance ratio to obtain a posterior PDF for each galaxy, which is then normalised. From this PDF we produce summary statistics assuming a skew-normal distribution, recording the mean (\texttt{logdist $= \langle\eta_n\rangle$}), standard deviation (\texttt{logdist\_err $= \sigma_{\eta_n}$}) and skew (\texttt{logdist\_alpha $ = \gamma_{\eta_n}$}) as the log-distance location, scale and shape parameters for each galaxy.

We compute log-distance ratios for all 108,810 targets that pass our initial selection criteria including non-primary observations (\texttt{primaryVdisp = False}) and Fundamental Plane outliers. These additional observations are useful for crossmatches to other DESI catalogues and for internal consistency checks. However, when doing population and cosmological analyses, only \texttt{primaryVdisp = True} targets should be used.

\subsubsection{Correction for group richness}\label{section:group_richness_correction}
When carrying out preliminary analyses of the clustering of the log-distance ratios in the FP catalogue (see companion papers \citealt{DESI_DR1_PV_dens_vel_corr, DESI_DR1_PV_power_spec, DESI_DR1_PV_maxlike}), an excess of clustering was identified in the velocity correlation function/power spectrum on small scales. We hypothesised this was due to a correlation between the richness of a group in which a FP galaxy resides and the measured log-distance ratio. A trend of this nature was found in \cite{Howlett_2022} using data from the SDSS. A correction was carried out in that work by fitting separate Fundamental Planes in bins of group richness, where it was found that the bias in the log-distance ratios could be removed by allowing the mean surface brightness $\bar{i}$ of the FP fit to vary between richness bins. The resultant fits exhibited a relatively smooth preference for lower mean surface brightnesses in richer groups, although at the time it was not clear whether this was due to intrinsic, physical, correlations between group environment and location of a galaxy on the FP, or data systematics such as difficulties de-blending photometry in dense clusters.

Unfortunately, there is not yet a group catalogue available for the DESI DR1 data, making testing and correcting for a similar systematic in this work more challenging. In lieu of this correction, a crude grouping algorithm was used to explore the relationship between the average log-distance ratio as a function of the number of other nearby FP measurements. Firstly, all pairs of galaxies in the full FP catalogue with a perpendicular and parallel separation (relative to the line of sight) of $r_{\perp}<1.75h^{-1}\mathrm{Mpc}$ and $r_{||}<8.5h^{-1}\mathrm{Mpc}$, were identified.\footnote{These are close to the scales on which the excess small-scale clustering was identified. Variations about these scales were investigated, and it was found that this choice gave close to the maximal number of independent groups, while the trend identified between group-size and average log-distance ratios was reasonably robust to variations in the pair-matching radii.} Then, these pairs were converted into $12,736$ distinct groups by identifying repeated indices (i.e., if a pair [A, B] and a pair [B, C] exist, these were grouped into a triplet [A, B, C] and so on). Accounting for only galaxies with \texttt{primaryVdisp = True}, $67,055$ of the FP galaxies were found to reside in groups of at least two members. The remaining $31,238$ were designated as field galaxies.

\begin{figure}[hbt!]
 \centering
 \includegraphics[width=\linewidth]{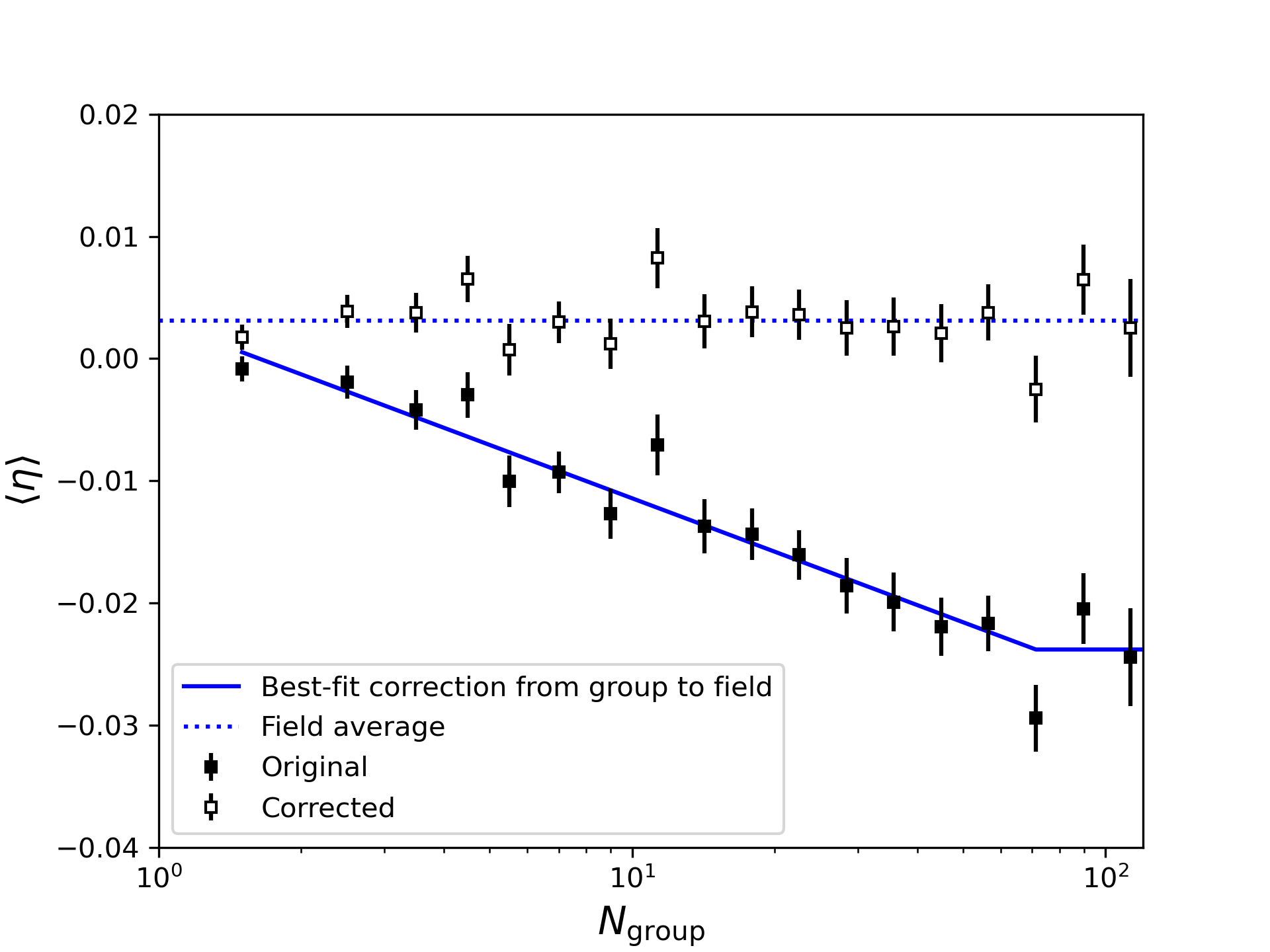}
 \caption{Average log-distance ratio as a function of group-size in the FP catalogue before (solid points) and after (open points) correction for this bias. The solid line indicates the piecewise fit used for the correction. The dotted horizontal line is the average log-distance ratio for field galaxies (i.e., those not in any groups).}
 
 \label{fig:FP_groupcorr}
\end{figure}

Using this grouping, a clear difference between the average log-distances of galaxies in groups and in the field, increasing with group size, was identified. This is shown in Fig.~\ref{fig:FP_groupcorr}. The direction of this trend --- larger groups having, on average, more negative log-distance ratios --- matches the trend seen in SDSS data \citep{Howlett_2022}, although its amplitude in DESI is roughly half that found in the SDSS. The fact that a similar trend is found in both independent datasets is intriguing and points perhaps more towards a physical origin for the correlation. On the other hand, the difference in amplitude works against this hypothesis. It should be noted, however, that a weaker signal may also arise from our ad-hoc method of identifying groups.

In any case, the log-distance ratios in the FP catalogue are corrected for this trend by fitting a piece-wise linear$+$constant function to the bias as a function of group-size relative to the field-average. Our best-fitting function is given by 
\begin{equation}
\langle \eta_{\mathrm{group}} \rangle - \langle \eta_{\mathrm{field}} \rangle = 
    \begin{cases}
        -0.0145\log_{10}(N_\mathrm{group}) & N_\mathrm{group} < 71 \\
        -0.0268 & \mathrm{otherwise}
    \end{cases},
\end{equation}
and gives a $\chi^{2}/\mathrm{dof} = 16.3/16$. This correction was applied to each individual galaxy based its group occupancy. Log-distance ratios both with and without this correction are included in the final FP catalogue, but the group-corrected distances were used for subsequent zero-point calibration \citep{DESI_DR1_PV_zp} and clustering measurements \cite{DESI_DR1_PV_dens_vel_corr, DESI_DR1_PV_power_spec, DESI_DR1_PV_maxlike}.

\subsubsection{Zero-Point Calibration and Peculiar velocities}\label{section:zp_and_pvs}
Initially, the zero-point of the Fundamental Plane relation is unanchored to independent distance measurements, but set under the assumption of zero bulk radial flow. In other words, on average, log-distance ratios are 0 with no redshift evolution. The physical zero-point is set as the last step of fitting the Fundamental Plane since it is a simple offset $\Delta\eta$ to shift the Hubble Diagram to that of any given independent distance indicators (the `calibrators').

The calibration of the zero-point for the DESI DR1 PV sample is done in the companion paper by \citet{DESI_DR1_PV_zp}. There, we perform a joint fit of FP and TF data and independent supernovae Ia distances from the Pantheon+SH0ES sample \citep{SH0ES_2022,Pantheon+}.
To do this, we make use of galaxy groups by equating the distance to any group member to all members of the group, then correct for any sample-wide average offset between DESI DR1 distances to those groups and the independently measured distances (the `zero-point' offset). This step is essential to improve the statistical uncertainty introduced by minimal direct overlap of SNe Ia in DR1 FP and TF galaxies. Once the zero-point of the FP and TF relations have been tied to independent distances, we can measure the expansion rate, $H_0$, directly from the DR1 distance-redshift relation. The method of finding the galaxy groups is described in \cite{DESI_DR1_PV_zp} along with the FP+TF global zero-pointing process and $H_0$ constraints. 

Log-distance ratios are a measure of the difference between observed and true galaxy sizes, which we attribute to peculiar velocities; peculiar velocities alter observed redshifts such that the distance we infer differs from the true distance. Estimating the true sizes/distances from the FP relation introduces significant scatter that is not related to peculiar velocities, but in principle $\eta$ is a direct inference of PV. The same is true for TF measurements, except in that case we measure distance moduli $\mu$ and compare to a fiducial model such that the PV measure is $\Delta\mu=-5\eta$.
To convert from $\eta$ to peculiar velocities $\hat{v}$, we use the \citet{Carreres_2023_pv_est} estimator
\begin{equation}\label{eq:CarreresPVs}
    \hat{v}(\eta) = c\ln{10}\left(\frac{(1+z_{\text{cmb}})c}{H(z_{\text{cmb}})d(z_{\text{cmb}})}-1\right)^{-1}\eta,
\end{equation}
which reduces to other common estimators used in peculiar velocity studies under different approximations. \footnote{Note that, in this process $H(z_{\rm CMB})$ and $d(z_{\rm CMB})$ are the redshift dependant Hubble parameter and comoving distance respectively, as computed from the observational CMB-frame redshifts.}

\citet{DESI_DR1_PV_zp}, uses the same fiducial cosmology as this work (flat $\Lambda$CDM model with $H_0 = 100$\kmsmpc and $\Omega_m = 0.3151$) during their zero-pointing and $H0$-fitting processes \red{which is later updated to the best-fit value $H_0 =73.7$\kmsmpc for computing peculiar velocities. This means that when applying the $\Delta\eta $ zero-point shift to the DESI DR1 FP data we need to take into consideration the difference between the fiducial and fitted models. We do so by taking the fiducial zero-point $\eta_\mathrm{zp} = -0.139 \pm 0.007$ \citep[Table 2,][]{DESI_DR1_PV_zp} and subtracting the log ratio of the $H_0$ values to obtain the final shift $\Delta\eta =-0.136 -\log\left(H_0^\mathrm{fit}/H_0^\mathrm{fid}\right) = -0.006 \pm 0.007$ which is applied to our data.}

\section{Fundamental Plane Results}\label{section:FP_results}
\begin{figure*}[hbt!]
 \centering
 \includegraphics[width=\linewidth]{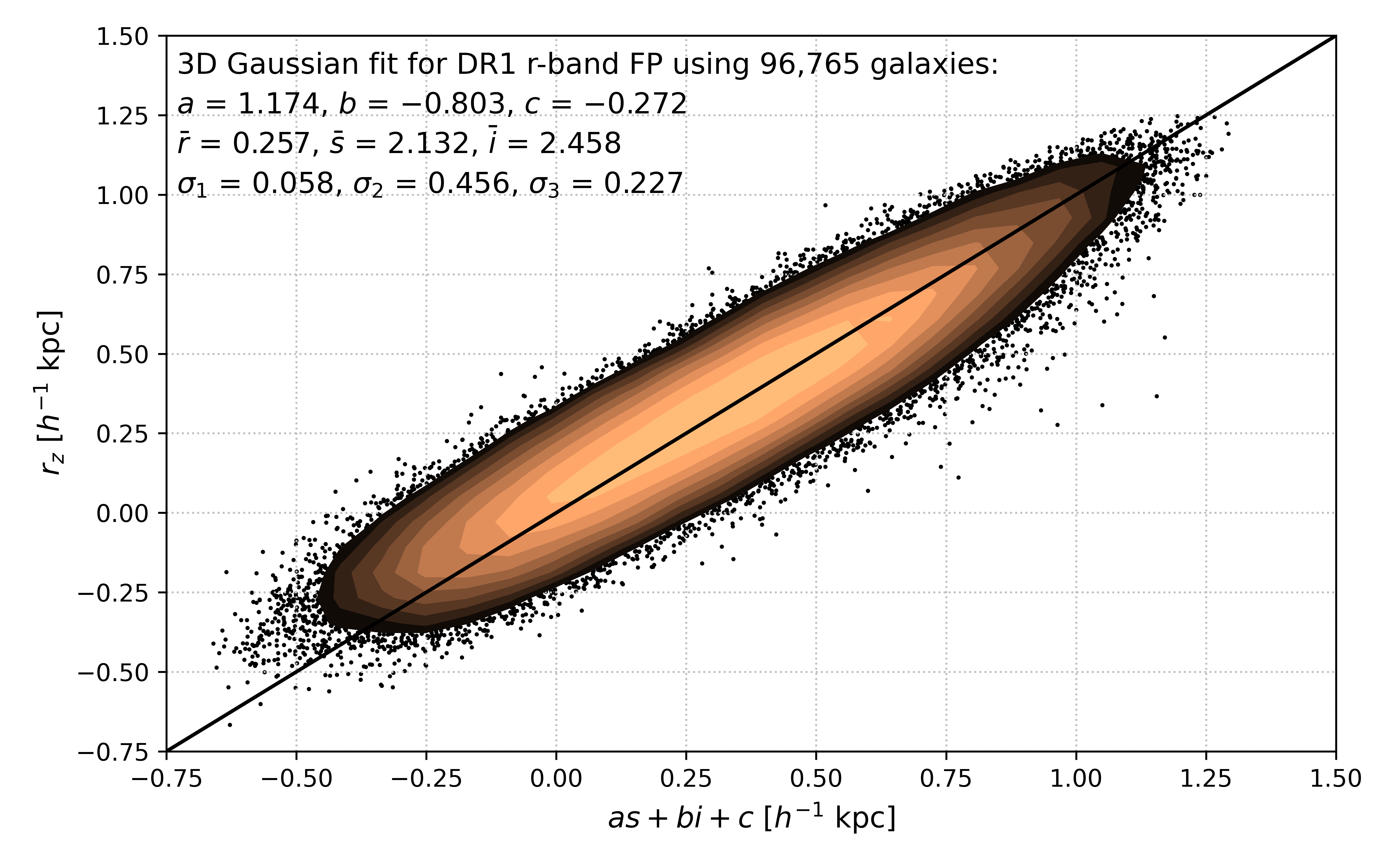}
 \caption{The projected Fundamental Plane for the DR1 \texttt{FPcalibration} sample. The figure shows the measured effective radii ($r_z$) as a function of the predicted effective radii ($r_t = as+bi+c$). The solid black line represents the one-to-one relation and the fiducial fitted Fundamental Plane parameters are superimposed.}
 \label{fig:fid_FP_contour}
\end{figure*}

A projection of our best-fitting Fundamental Plane for the DR1 Fundamental Plane calibration sample (\texttt{FPcalibration = True}) is shown in Figure \ref{fig:fid_FP_contour}. This plot illustrates the relation between the effective radii deduced from our fiducial cosmology ($r_z$) and the effective radii predicted by our 3D Gaussian model ($r_t = as+bi+c$), and is accompanied by the best-fit Fundamental Plane parameters. Vertical offsets in the y-direction from the one-to-one line (black) roughly correspond to the measured peculiar velocity for each galaxy.\footnote{This is not a direct calculation; the procedure for the derivation of the individual log-distance ratios and peculiar velocities using the full likelihood function is described in sections \ref{section:logdist_fit} and \ref{section:zp_and_pvs}} The contour is slightly tilted when compared to the one-to-one line. This is expected as we up-weight faint galaxies during the fitting process to account for missing galaxies at high redshift due to the selection function. These up-weighted galaxies reside in the lower left part of the contour below the one-to-one line, resulting in the apparent overabundance of galaxies above the one-to-one line with high surface brightness which appears as the slight tilt compared to the one-to-one relation. 

\begin{figure*}[hbt!]
 \centering
 \includegraphics[width=\linewidth]{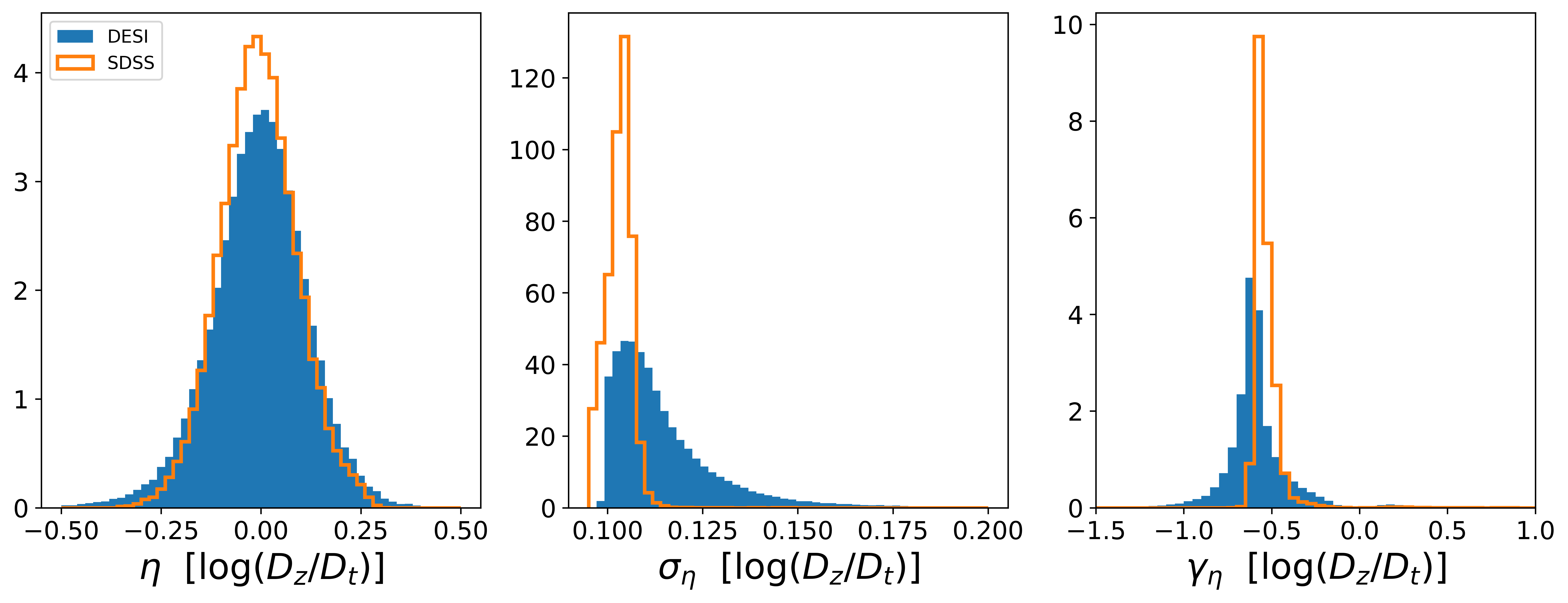}
 \caption{Distribution across all galaxies in the DR1 Fundamental Plane sample (blue) of the log-distance likelihood distribution parameters. The left panel shows the distribution of measured log-distance ratios (mean of individual galaxy's log-distance likelihood distribution). The middle panel shows the distribution of measured log-distance errors (standard deviations) and the right panel shows the distribution of skews. For comparison we show the SDSS distributions (orange)\red{. The mean of the log-distances are consistent between the two surveys but DESI has larger log-distance errors and skews. These discrepancies are attributed to larger velocity dispersion errors and larger intrinsic scatter in the DESI FP.}}
 \label{fig:fid_logdist_hists}
\end{figure*}

Figure \ref{fig:fid_logdist_hists} depicts the distributions of the mean log-distance ratio, $\eta$, log-distance ratio error, $\sigma_\eta$, and log-distance ratio skewness, $\gamma_\eta$, measurements across the \texttt{primaryVdisp} DR1 sample. The mean log-distance and log-distance error measurements have similar distributions to the SDSS sample but with slightly higher spread in the mean log-distance measurements, $\langle\sigma_\eta^\mathrm{DESI}\rangle = 0.12$ dex for DESI and $\langle\sigma_\eta^\mathrm{SDSS}\rangle = 0.10$ dex for SDSS. This translates to a distance error of $26\%$ in the DESI sample compared to $23\%$ in the SDSS sample. 

One method of quantifying the intrinsic scatter in the Fundamental Plane is using the r-direction scatter, $\sigma_{r,\mathrm{int}}^2$, from the FP model covariance matrix \citep[Equation 3B]{Howlett_2022}. Using the fiducial fits, DESI has a larger intrinsic $r$-direction scatter in the FP relation at $\sigma_{r,\mathrm{int}}^2 = 0.097\pm 0.002$, compared with $\sigma_{r,\mathrm{int}}^2 = 0.0610$ for SDSS. This could be due to several causes. SDSS, unlike DESI, underwent a visual inspection process to remove contaminant spiral galaxies. Due to the much larger size of the DESI sample this was not conducted and there is likely to be a non-negligible number of contaminants remaining in the sample. As early-type and star-forming galaxies typically occupy different FP-like relations, this would increase the intrinsic scatter of the calibration. Additionally, DESI has larger $r$-parameter measurement errors than SDSS which may be contributing to the larger intrinsic error in the calibration. 

The DESI distribution of velocity dispersion errors ($s$-parameter) is heavily skewed with a long tail of high errors. As a result, the mean velocity dispersion error is $\epsilon_s^\mathrm{DESI} = 0.0457$~dex but the peak of the distribution lies at approximately $0.025$ dex. This is much closer to, but still almost double, the mean SDSS velocity dispersion error $\epsilon_s^\mathrm{SDSS} = 0.0147$. Additionally, the distribution of the DESI velocity dispersion errors has a dispersion 5 times larger than SDSS. Neither of these discrepancies is meaningfully reduced by adopting the SDSS aperture correction or by removing the error scaling introduced in Section \ref{section:vdisp_internal_consistancy}. We intend to investigate the reasons behind why DESI velocity dispersions are so large and dispersed as part of a systematics review using DR2.

In contrast, we find that DESI has smaller photometric errors than SDSS. For surface brightness, we measure a fiducial model mean measurement error of $\epsilon_i^\mathrm{DESI}=0.012$ dex which is comparable to $\epsilon_i^\mathrm{SDSS}=0.014$ dex for SDSS. However, this measurement is substantially reduced to $\epsilon_i^\mathrm{DESI}=7.75\times10^{-4}$ dex if we adopt the SDSS evolution correction value of $Q=0.85$, indicating that the primary error source in this parameter is due to the uncertainty in this correction, which was not propagated in the SDSS analysis. Similarly, the mean $r$-parameter error error is an order of magnitude smaller in DESI when compared to SDSS; $\epsilon_r^\mathrm{DESI} = 6.13\times 10^{-4}$ dex vs $\epsilon_r^\mathrm{SDSS} = 6.98\times 10^{-3}$ dex.  

\begin{figure}[hbt!]
 \centering
 \includegraphics[width=\linewidth]{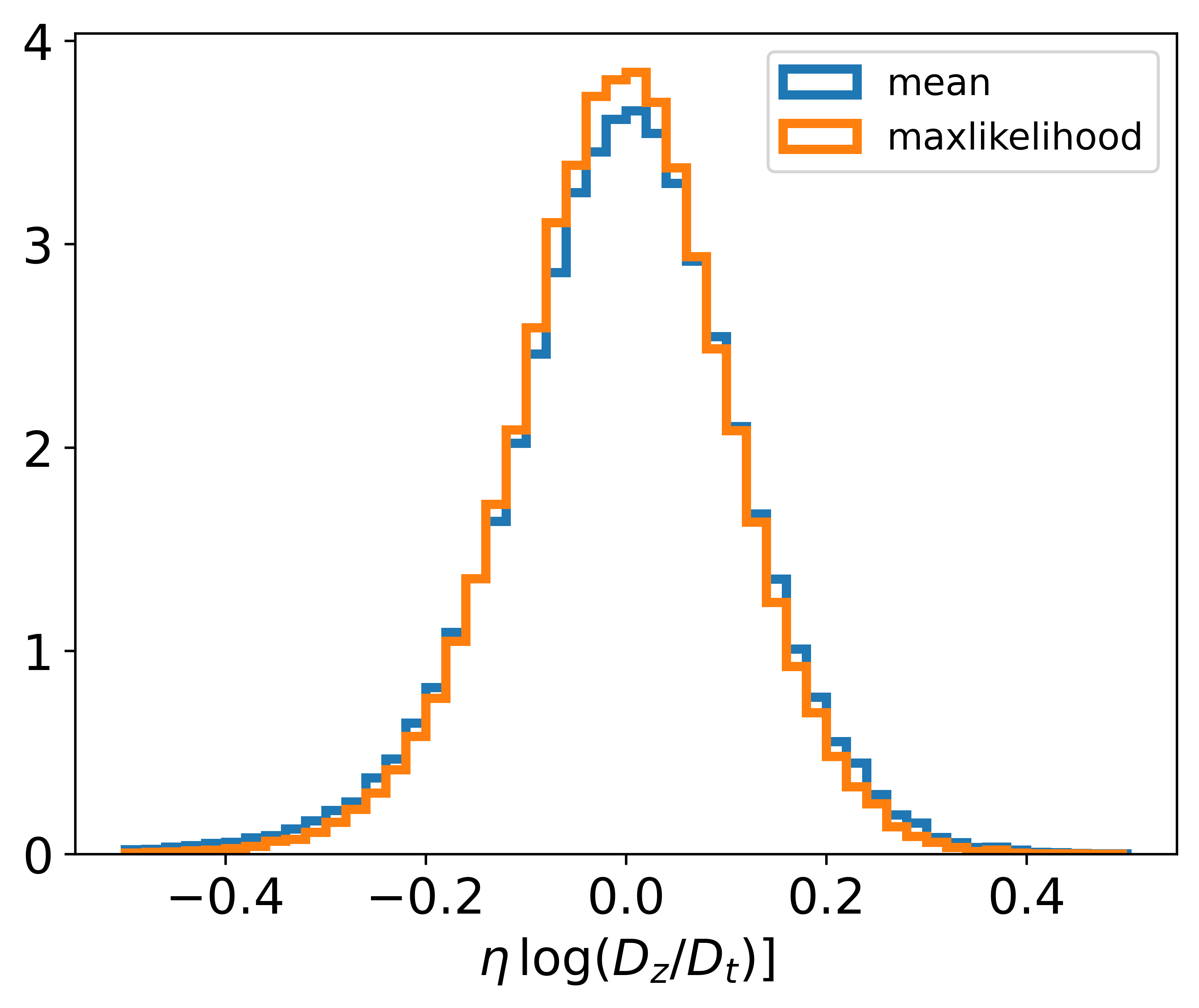}
 \caption{Distribution of galaxy log-distances when taking mean (blue; fiducial method) vs maximum likelihood (orange) of the individual galaxy log-distance pdf. A relative zero-point has been applied to approximate the result of repeating Section \ref{section:zp_and_pvs} using the maximum likelihood values. \red{There is minimal discrepancy between the two distributions, as such, using the `mean' measure over the `maximum likelihood' should not have a noticeable effect on the reported distances.}}
 \label{fig:mean_vs_maxlikelihood}
\end{figure}

Both the SDSS and DESI DR1 samples prefer negatively skewed log-distance PDFs (right panel, Figure \ref{fig:fid_logdist_hists}). This indicates that for the majority of galaxies in our sample the most likely log-distance ratio is slightly larger than the mean of that galaxy's log-distance PDF. The PV estimator we are using in this work assumes that the galaxy log-distance PDFs are Gaussian. As such, we present galaxy distances and PVs calculated using the `mean' log-distances, as these measurements better represent the underlying PDFs when operating under this assumption. However, it is important to understand any differences that arise in the sample from using the `maximum likelihood' log-distance values instead.

In Figure \ref{fig:mean_vs_maxlikelihood}, we present a comparison between the log-distance sample distribution using the `mean' (blue) and `maximum likelihood' (orange) values for the galaxies' log-distances. When creating this plot, we apply a relative zero-point of $\eta_{mean} - \eta_{ML} = -0.10$ to the `maximum likelihood' log-distances, replicating the effects of repeating an external zero-point (Section \ref{section:zp_and_pvs}) on the 'maximum likelihood' log-distance measurements. The two distributions are very similar with only a slight increase in the dispersion ($\sigma = 0.124$ vs $\sigma =0.116$) when using the `mean' over the `maximum likelihood' log-distance values. From this we conclude that using the `mean' measure for log-distance should not have a noticeable effect on our final reported distances for the sample as a whole when compared to using the `maximum likelihood' value. However, we caution that if using the distances for individual galaxies, the `mean' and `maximum likelihood' values for log-distance vary by up to $0.56$ dex in the most highly skewed cases.

\begin{table*}[t]%[hbt!]
 \centering
 \begin{tabular*}{\textwidth}{@{\extracolsep{\fill}} l *{10}{l}}
\toprule
 & $N_{gal}$ &  a&  b&  $\bar{r}$&  $\bar{s}$&  $\bar{i}$&  $\sigma_1$&  $\sigma_2$& $\sigma_3$&$\sigma_{r,\mathrm{int}}^2$\\
\midrule
  & &1.17&-0.803&0.257&2.132&2.458&0.0582&0.456&0.227&0.097\\
\multirow{-2}{*}{Fiducial}& \multirow{-2}{*}{96,758} &$\pm$0.01&$\pm$0.005&$\pm$0.008&$\pm$0.003&$\pm$0.003&$\pm$0.0002&$\pm$0.003&$\pm$0.003&$\pm$0.002 \\
\midrule
No sigma clipping&98,292&\textbf{1.23}&\textbf{-0.787}&0.256&2.124&\textbf{2.443}&\textbf{0.0630}&\textbf{0.475}&\textbf{0.248}&\textbf{0.106}\\
\midrule
$\sigma_{vdisp} <360\kms$&96,745&1.18&-0.803&0.256&2.131&2.457&0.0581&0.457&0.224&0.097\\
\midrule
$\sigma_{vdisp} <500\kms$&96,770&1.18&-0.803&0.256&2.131&2.457&0.0581&0.456&0.224&0.097\\
\midrule
$\sigma_{vdisp} >60\kms$&95,153&1.17&-0.803&0.267&\textbf{2.145}&2.464&0.0581&0.459&\textbf{0.206}&0.096\\
\midrule
$\sigma_{vdisp} >46\kms$&97,199&1.19&-0.803&0.251&2.125&2.454&0.0581&0.456&0.233&0.098\\
\midrule
$m_r < 17$ mag&69,348&1.17&-0.805&0.246&\textbf{2.113}&\textbf{2.440}&\textbf{0.0598}&0.453&0.236&0.097\\
\midrule
$m_r > 14$ mag&94,250&1.18&-0.801&0.250&2.128&2.460&0.0581&0.455&0.223&0.096\\
\midrule
H$\alpha$ EW $<10$ \AA&88,733&1.19&-0.798&0.256&\textbf{2.146}&\textbf{2.471}&\textbf{0.0566}&0.452&0.224&0.095\\
\midrule
H$\alpha$ EW $<1$ \AA&66,240&\textbf{1.22}&\textbf{-0.783}&\textbf{0.228}&\textbf{2.156}&\textbf{2.507}&\textbf{0.0533}&\textbf{0.439}&0.227&\textbf{0.090}\\
\midrule
$\Delta\chi_z^2>50$&96,767&1.18&-0.803&0.256&2.131&2.457&0.0581&0.457&0.224&0.097\\
\midrule
no $\Delta\chi_z^2$ cut&96,770&1.18&-0.803&0.256&2.131&2.457&0.0581&0.456&0.224&0.097\\
\midrule
$n_s>4$&59,281&1.19&-0.799&\textbf{0.336}&\textbf{2.177}&\textbf{2.400}&0.0584&\textbf{0.446}&\textbf{0.204}&\textbf{0.090}\\
\midrule
$B/A >0.5$&76,276&1.19&\textbf{-0.823}&\textbf{0.285}&2.130&\textbf{2.411}&\textbf{0.0608}&\textbf{0.440}&0.229&0.094\\
\midrule
$Q = 1.5$&96,771&1.17&-0.802&0.256&2.131&\textbf{2.446}&0.0582&0.457&0.223&0.097\\
\midrule
$Q = 0.7$&96,772&1.19&-0.803&0.256&2.131&\textbf{2.468}&0.0585&0.456&0.224&0.097\\
\midrule
$Q = 0.85$ (SDSS value) &96,770&1.19&-0.803&0.256&2.131&2.464&0.0585&0.456&0.224&0.097\\
\midrule
$\alpha = -0.09$&96,720&1.21&\textbf{-0.827}&0.256&\textbf{2.145}&2.457&\textbf{0.0595}&0.456&0.222&0.099\\
\midrule
$\alpha = -0.03$&96,817&1.15&\textbf{-0.780}&0.256&\textbf{2.116}&2.455&0.0588&0.457&0.228&0.095\\
\midrule
$\alpha = -0.04$ (SDSS value)&96,797&1.16&-0.788&0.256&\textbf{2.121}&2.456&\textbf{0.0589}&0.457&0.226&0.095\\
\bottomrule
\end{tabular*}
\caption{Fundamental Plane parameters for the DESI DR1 sample and variations. In each row we make a single change to our selection criteria, method, or parameter choice, with each change described in the left column. The uncertainties on the fiducial Fundamental Plane parameters are derived from the scatter seen in mock realisations of our data (Bautista et al., in prep), and values that lie more than $3\sigma$ from our fiducial parameters have been \textbf{bolded}.}
\label{table:FP_fit_results}
\end{table*}

\subsection{Internal Validation}\label{section:IntValidation}

The Fundamental Plane is a population-calibrated relation, meaning it is heavily dependent on the properties of the sample to which it is fit and can vary significantly between different populations of galaxies. In this work, we have attempted to isolate our sample to luminous red early-type galaxies using strict selection criteria as described in sections \ref{selection:primary} and \ref{selection:vdisp_cuts}. Some of these cuts, while important, are known to introduce biases, such as Malmquist bias from the magnitude and redshift cuts. These biases have been accounted for to the best of our ability in the likelihood function $f_n$ and $S_n$ terms (Equation \ref{eq:FP_likelihood}). However, it is important that we test our methodology for additional sources of biases by exploring our fiducial analysis choices for how they affect the recovered Fundamental Plane relation and log-distance ratios.

We looked at two primary factors when deciding on our final fiducial selection criteria: 1) minimising any tilt in the relationship between redshift and log-distance, 2) the trade-off between minimising the intrinsic scatter and maximising the sample size. The first of these criteria arises from our underlying assumption that the recovered peculiar velocities should not depend strongly on redshift. We evaluate this by plotting the mean log-distances, with uncertainties in redshift bins. The second criterion is more subjective and as such, unless there was a clear and significant improvement in the dispersion $\sigma_r^2$, we maintained the selection criteria used in target selection \citep{DESI_PV_target_selection} and the DESI EDR Fundamental Plane \citep{Said_EDR}. 

In addition, we wanted to identify possible weaknesses in our selection criteria for further analysis in DR2. We tested a variety of samples that fell within the targeting criteria, including modifications to velocity dispersion and $\Delta \chi^2_z$ cuts, tighter constraints on magnitude, S\'{e}rsic index and $b/a$ ratio as well as introducing a cut on the H$\alpha$ equivalent width. We also tested removal of the sigma-clipping step in our method as well as using a variety of evolution correction and aperture correction values from the literature to compute our Fundamental Plane input parameters. The Fundamental Plane parameters for each of these fits, including the resulting sample size and r-direction intrinsic scatter, $\sigma_r^2$, are presented in Table \ref{table:FP_fit_results}.

Variations to the $b/a$ ratio, upper and lower velocity dispersion cuts and $\Delta\chi^2$ cuts had no or minimal effect on the resulting log-distance ratios and less than $5\%$ variation in $\sigma_r^2$. Decreasing the maximum magnitude cut shifted the mean log-distance higher but did so in a very uniform way across the sample, as such this shift would be removed during the zero-point process, resulting in equivalent distances and peculiar velocities in the fiducial sample. Additionally, while decreasing the magnitude limit slightly improves the surface brightness error, it significantly decreases the total sample size and increases the intrinsic $\sigma_r^2$ scatter, having the opposite effect to our goal.

If we instead tighten our lower magnitude cut to $m_r > 14$ mag, we observe that the tilt in the log-distance redshift relation flattens out earlier than the fiducial model at $z\sim0.025$ compared to $z\sim0.04$. This cut also results in a minimal decrease in the overall sample size (loss of only $\sim2500$ galaxies) which is ideal at first glance. However, these removed galaxies lie almost entirely at low redshifts with the cut removing $\sim70\%$ of the galaxies with $z<0.02$ and $\sim25\%$ of galaxies with $z<0.04$.  This is the primary redshift domain used for our zero-point calibration and as a result we reduce our total number of calibrators by $35\%$ . Additionally, as FP uncertainties scale with redshift, this is also the domain on which we have the tightest constraint on galaxy distances and PVs. As such, we do not want to remove such a large percentage of these galaxies for a minimal decrease in the intrinsic scatter.

Increasing the S\'{e}rsic index cut or implementing a hydrogen alpha equivalent width (H$\alpha$ EW) cut both slightly reduce the $\sigma_r^2$ scatter but significantly reduce the sample size and exacerbate the redshift-log-distance tilt. It is important to note that the S\'{e}rsic index cut, and the H$\alpha$ EW cut have the greatest improvement on the $\sigma_r^2$ scatter out of the tested samples. As these parameters are both indicators of star formation, it indicates that there is likely some small residual spiral contamination in our sample. We have visually identified some of these contaminants and will investigate further ways to clean the sample in the DR2 analysis based on comprehensive visual inspection of the accumulated data.

\subsubsection{Evolution and aperture corrections} \label{section:evo_correction}
\begin{figure}[t]
  \includegraphics[width=\linewidth]{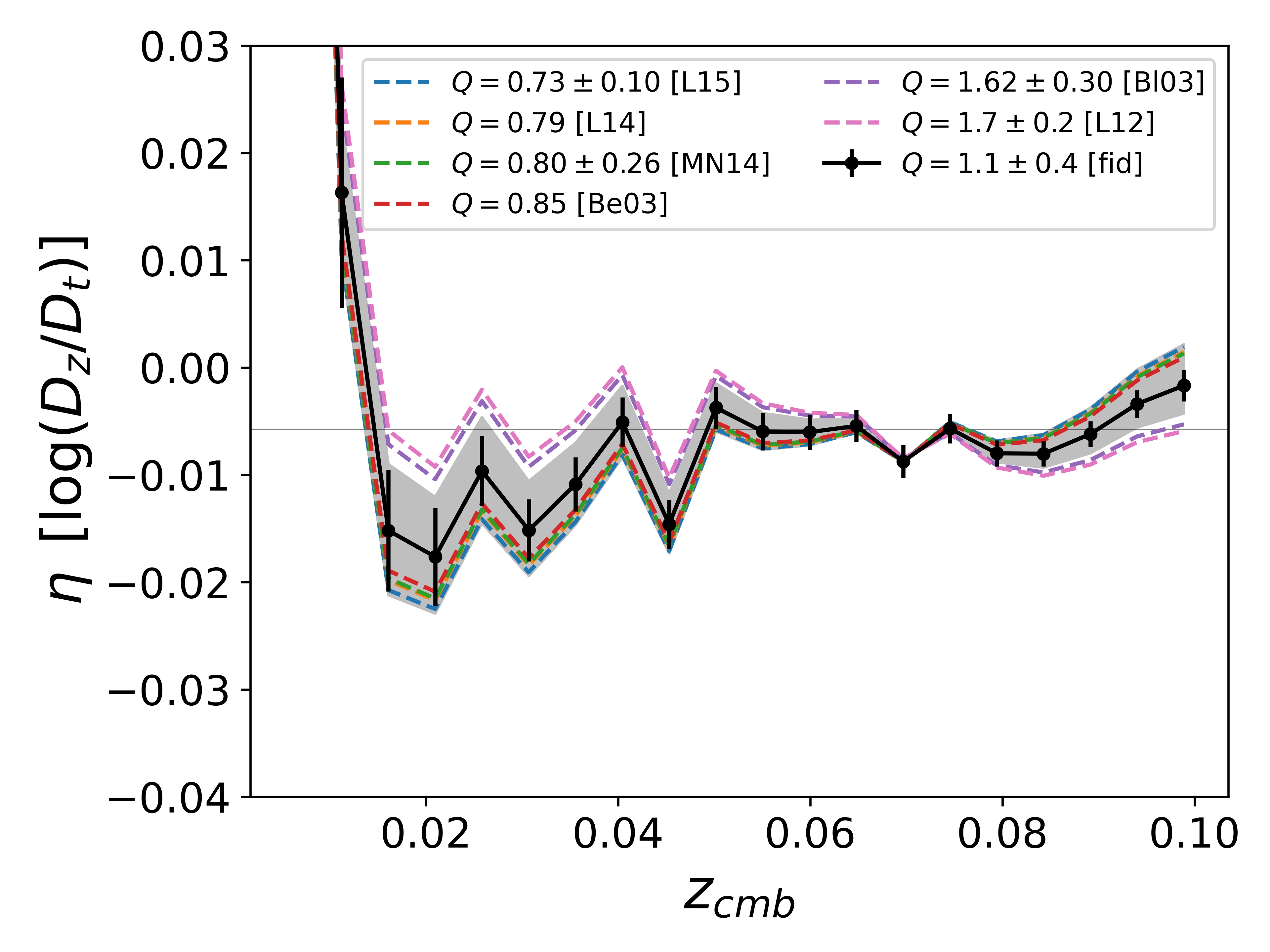}
    \captionof{figure}{Mean log-distance ratio in redshift bins for Fundamental Plane fits using the Fiducial aperture correction $\alpha_\mathrm{fid} = -0.06\pm 0.03$ while varying the evolution correction parameter. In the solid black we show our fiducial sample with the shaded region indicating the spread between $1\sigma$ variations of our fiducial evolution correction. For comparison the Fundamental Plane realisations using the literature evolution correction values are plotted in the coloured dashed lines. From this plot we can see that the chosen fiducial model is close to flat between $0.04<z<0.09$, bringing it into agreement with the assumption of redshift-independent peculiar velocities.}
    \label{fig:evocorr_logdist}
\end{figure}

\begin{figure}[t]
    \includegraphics[width=\linewidth]{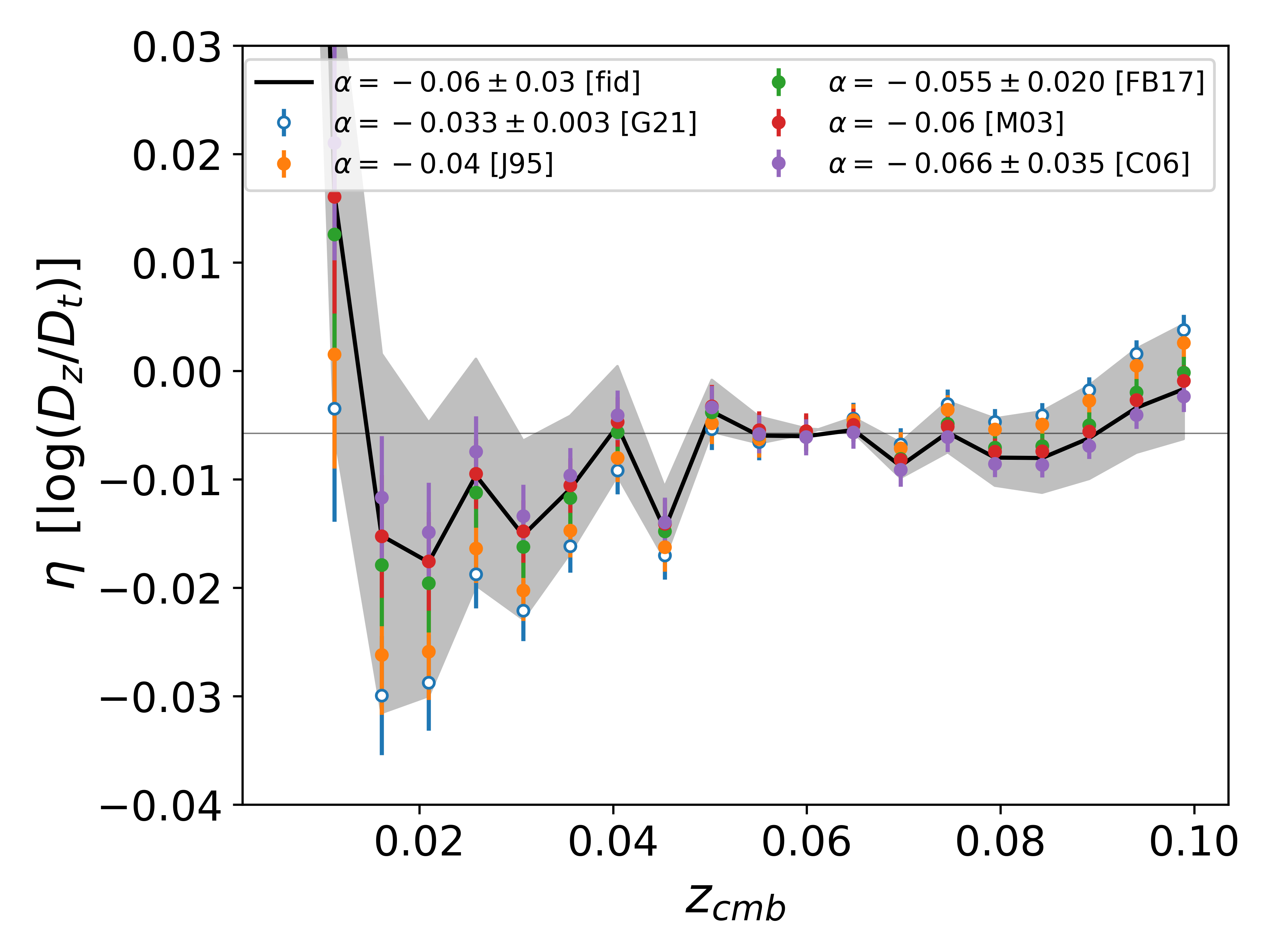}
    \captionof{figure}{\red{As in Figure \ref{fig:evocorr_logdist}, but with fixed evolution correction $Q_\mathrm{fid} = 1.1\pm 0.4$ and varying aperture correction values from prior works. The \citeauthor{Jorgensen_1995} (J04, orange), \citeauthor{Falcon-Barroso_2017} (FB17, green), \citeauthor{Mehlert_2003} (M03, red) and \citeauthor{Cappellari_2006} (C06, purple) values are all measured using samples containing only early-type galaxies, while the \citeauthor{deGraaff_2021} value (G21, Blue) was measured from a mixed sample. As with Figure \ref{fig:evocorr_logdist}, the chosen fiducial aperture correction model is consistent with the assumption of redshift-independent peculiar velocities and has sufficiently large error to encompass variation arising from the different correction values.}}
    \label{fig:sigma_corr_logdist}
\end{figure}

Two parameters that were found to effect the log-distance redshift tilt are the two correction parameters; the aperture correction parameter, $\alpha$, and the evolution correction parameter, $Q$. To reiterate, the aperture correction parameter enters into the Fundamental Plane as a redshift-independent correction to the $s$ input parameter (Equation \ref{eq:FP_s}). It corrects for the fact that the proportion of the galaxy being captured by the spectrograph varies with the on-sky radius of the galaxy (in  \arcsec). This has the strongest effect of our two `corrections', with more negative aperture correction indices rotating the log-distance-redshift plane clockwise around $z\sim0.06$. The evolution correction parameter --- a redshift-dependent term that corrects for galaxies dimming slightly as they evolve and is incorporated into the calculation of the Fundamental Plane input parameter $i$ --- has a smaller effect, although similar in nature to the aperture correction, rotating the mean log-distance ratios clockwise around $z\sim0.07$ for larger values of $Q$. During our analysis, we attempted to fit for $Q$, both as a Fundamental Plane parameter and separately following the method outlined by \citet{Saulder_2013}. Unfortunately, we were unable to converge on a value using these methods due to degeneracies with other parameters and analysis choices. Instead, we incorporated the wide spread of literature values for both the aperture correction and evolution correction terms into our fiducial analysis. Due to the similar effect these parameters had on the measured log-distances, we chose our fiducial parameters as $\alpha_\mathrm{fid} = -0.06\pm 0.03$, $Q_\mathrm{fid} = 1.1\pm0.04$, the pair that best flattened the log-distance-redshift tilt while staying within the published literature measurements.

In the literature, $Q$ is often included in works that fit a linearly evolving luminosity function to galaxy populations spanning a wide redshift range. We identified five papers that measured the $r$-band luminosity function for early-type galaxies and fit a single population $Q$ correction. They measure values as follows: $Q_{Bl03} = 1.62 \pm 0.3$ \citep{Blanton_2003}, $Q_{L12} = 1.7\pm 0.2$ \citep{Loveday_2012}, $Q_{L14} = 0.79$ \citep{Loveday_2014}, $Q_{MN14} = 0.80\pm0.26$ \citep{McNaught_2014}, and $Q_{L15} = 0.73 \pm 0.1$ \citep{Loveday_2015}. We also include the \cite{Bernardi_2003} value $Q_{Be03} = 0.85$ that was measured using a modified Fundamental Plane relation. This is the value that was used by both the SDSS and DESI EDR Fundamental Plane surveys \citep{Howlett_2022, Said_EDR}. Our fiducial evolution correction value comes from the mean of the previously listed literature values, taking the standard deviation as error, giving us $Q_\mathrm{fid} = 1.1\pm 0.4$. In Figure \ref{fig:evocorr_logdist}, we plot the log-distance ratios resulting from the Fundamental Plane fits for each of the aforementioned $Q$ values and their respective uncertainties, while fixing the aperture correction at the fiducial value. In the solid black line, we show the fiducial mean log-distance ratios with error bars representing the $1\sigma$ measurement uncertainties within each of the redshift bins, $\text{std}(\eta)/\sqrt{N_{bin}}$. Over-plotted in grey we show the range of the mean log-distances spanned by varying $Q_\mathrm{fid}$ by $1\sigma$ between $Q=0.7$ and $Q=1.5$. This demonstrates that our choice of fiducial evolution correction and adopting/propagating through a suitable uncertainty on $Q$ spans the results we would have obtained using any of the aforementioned literature values.

For the aperture correction, we found the following values in the literature: $\alpha_{G21} = -0.033 \pm 0.003$ \citep{deGraaff_2021}, $\alpha_{J95}=-0.04$ \citep{Jorgensen_1995}\footnote{This is the aperture correction value used previously by the SDSS and DESI EDR Fundamental Plane surveys.}, $\alpha_{FB17} = -0.055\pm 0.02$ \citep{Falcon-Barroso_2017}, $\alpha_{M03} = -0.06$ \citep{Mehlert_2003} and $\alpha_{C06} = -0.066 \pm 0.035$ \citep{Cappellari_2006}. For each measurement, we fit the Fundamental Plane and plot the resulting log-distances in Figure \ref{fig:sigma_corr_logdist}. All the measurements we tested, except for the \citet{deGraaff_2021} measurement, $\alpha_{G21}$, which was taken on a combined sample, were measured using samples that contained only early-type galaxies. \citet{Zhu_2023} also provide a range of aperture corrections that depend on S\'{e}rsic index, absolute $r$-band magnitude and $g-i$ colour, but this work was not used in our analysis as their range of binned measurements was found to have poor overlap with our DR1 sample. For our fiducial value, we took $\alpha_\mathrm{fid} = -0.06 \pm 0.03$ as when combined with our choice of evolution correction $Q_\mathrm{fid}= 1.1\pm 0.4$ the log-distance-redshift tilt is close to flat through the body of our sample $0.04<z<0.09$ (black line, Figure \ref{fig:evocorr_logdist}),  and the error bar on this correction ($\pm 0.03$) spans the range of measurements found in the literature.\label{section:internal_effect_alpha}

To summarise our internal consistency checks, the log-distance ratio measurements resulting from our Fundamental Plane fit are not heavily impacted by our choice of selection criteria. The selection cuts that have the largest reduction in intrinsic scatter are the S\'{e}rsic index and the hydrogen alpha equivalent width (H$\alpha$ EW) cuts. Both of these parameters are used to identify late-type star-forming galaxies  which indicates that there is some small spiral contamination in our sample. Due to the large reduction in sample size and relatively little gain for the cost, neither of these cuts has been implemented in the DR1 analysis, but we plan to further investigate this contamination and possible solutions in DR2. The two parameters that have the biggest effect on the log-distance ratio measurements are the evolution correction and aperture correction parameters. Varying either of these corrections within $1\sigma$ of the fiducial model has minimal effect on the recovered log-distance ratios of individual galaxies; the evolution correction giving $\eta\pm 0.002$ dex and the aperture correction giving $\eta \pm 0.005$ (typical log-distance ratio errors are $0.12$ dex). It should be noted that this is a redshift-dependent effect that more heavily impacts nearby galaxies, with $\Delta \eta \approx 0.05$ for the closest galaxies in our sample. As such, fixing these two parameters to a particular choice can affect results --- such as the bulk flow --- derived from a Fundamental Plane catalogue, highlighting the importance of marginalising over our uncertainty in these parameters as we have done in this work.

\subsection{\red{Comparison to CosmicFlows-4}}
\begin{figure}[t]
    \centering
    \includegraphics[width=\linewidth]{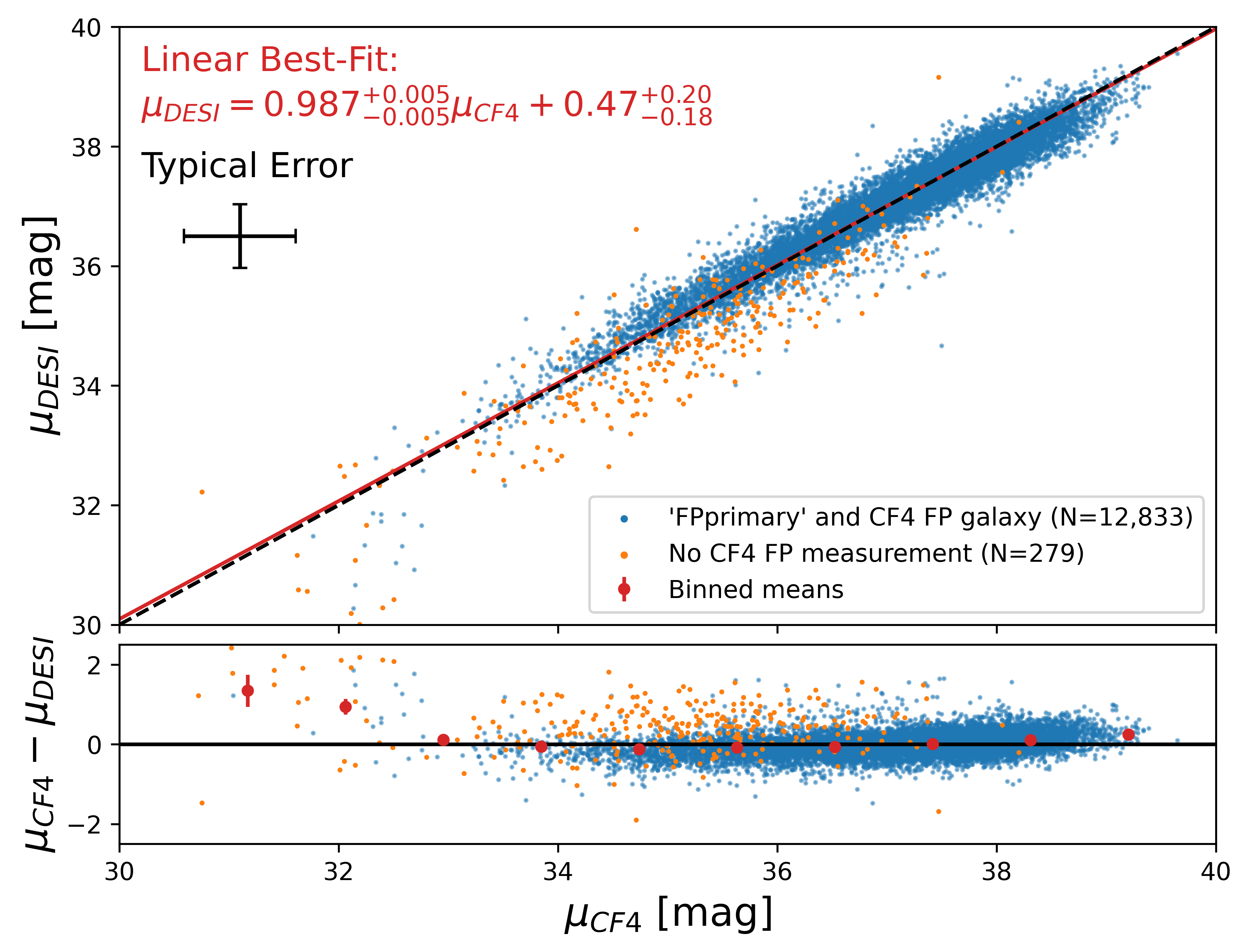}
    \caption{\red{Comparison of distance modulus as measured by CosmicFlows-4 \citep[weighted average across tracers,][]{cosmicflows-4} and DESI (this work). Overall, we find decent agreement between the two samples. Of the $13,112$ crossmatches, $279$ galaxies did not have a FP measurement in CF4, these have been highlighted in orange and account for the majority of outliers.}}
    \label{fig:CF4_dist_comp}
\end{figure}

\red{
It is important to compare our final distance measurements to external datasets in order to verify our results are free from bias. In this work we compare our results to the complete CosmicFlows-4 catalogue (CF4) as it is the most comprehensive distance catalogue currently available. We crossmatch to this catalogue by taking all matches within 5'' of each \texttt{primaryVdisp = True} DR1 FP galaxy. This results in a total of $13,112$ galaxies with measurements in both catalogues.

Our work presents distances in the form of log-distance ratios, $\eta$, which are cosmology independent. This differs from CF4 who report the distance modulus. By choosing a cosmological model it is possible to convert from log-distance ratio to distance modulus using
\begin{equation}
    \mu = 5 \log(d_{L}( z_{CMB} )) - 5\eta + 25 ,
\end{equation}
where $d_{L}( z_{CMB})$ is the luminosity distance of the galaxy in Mpc. For the DESI results we use the group and zero-point corrected log-distance ratios, Flat $\Lambda$CDM with $\Omega_m = 0.3151$ and $H_0 = 73.7$\kmsmpc as fit by \citet{DESI_DR1_PV_zp}.

We present the comparison between DESI DR1 FP and CF4 in Figure \ref{fig:CF4_dist_comp}. For this comparison we use the weighted average distance modulus across all tracers. We compute a best fit relation $\mu_{DESI} = 0.987^{+0.005}_{-0.005}\mu_{CF4} + 0.48^{+0.20}_{-0.18}$ which indicates good agreement between the samples. In orange we highlight the $279$ CF4 galaxies that have distance measurements but not via the Fundamental Plane. These galaxies account for the majority of the outliers in the sample --- given the almost constant offset between the orange and blue points in Figure \ref{fig:CF4_dist_comp} this could be indicative of either an internal zero-point difference between FP and non-FP measurements in the CF4 data, but it could also be that these galaxies are contaminants in our sample that should not be on the FP and so have slightly biased distances. It should also be noted that the CF4 and DESI measurements diverge for galaxies $\mu\lesssim33$ mag, which corresponds to redshift $z\sim 0.01$. These galaxies are removed from our clustering sample before computing any cosmology fits (see section \ref{section:clustering_cat}). The reason for these small number of disagreements are currently unknown but will be investigated further in DR2.
}

\section{Peculiar Velocity Catalogue}\label{section:pv_cat}

We present Fundamental Plane distances for $98,292$ galaxies measured from the DESI DR1 data.  A detailed description of the data format is given in the Data Availability section below, but to summarise, we include redshifts, sky locations, photometry, velocity dispersion parameters, fundamental plane parameters, log distance ratios, and the peculiar velocities we derive from these values. We also supply uncertainties on each measurement and flags to identify different sub-samples.

\begin{figure*}[tbp!]
 \centering
 \includegraphics[width=\linewidth]{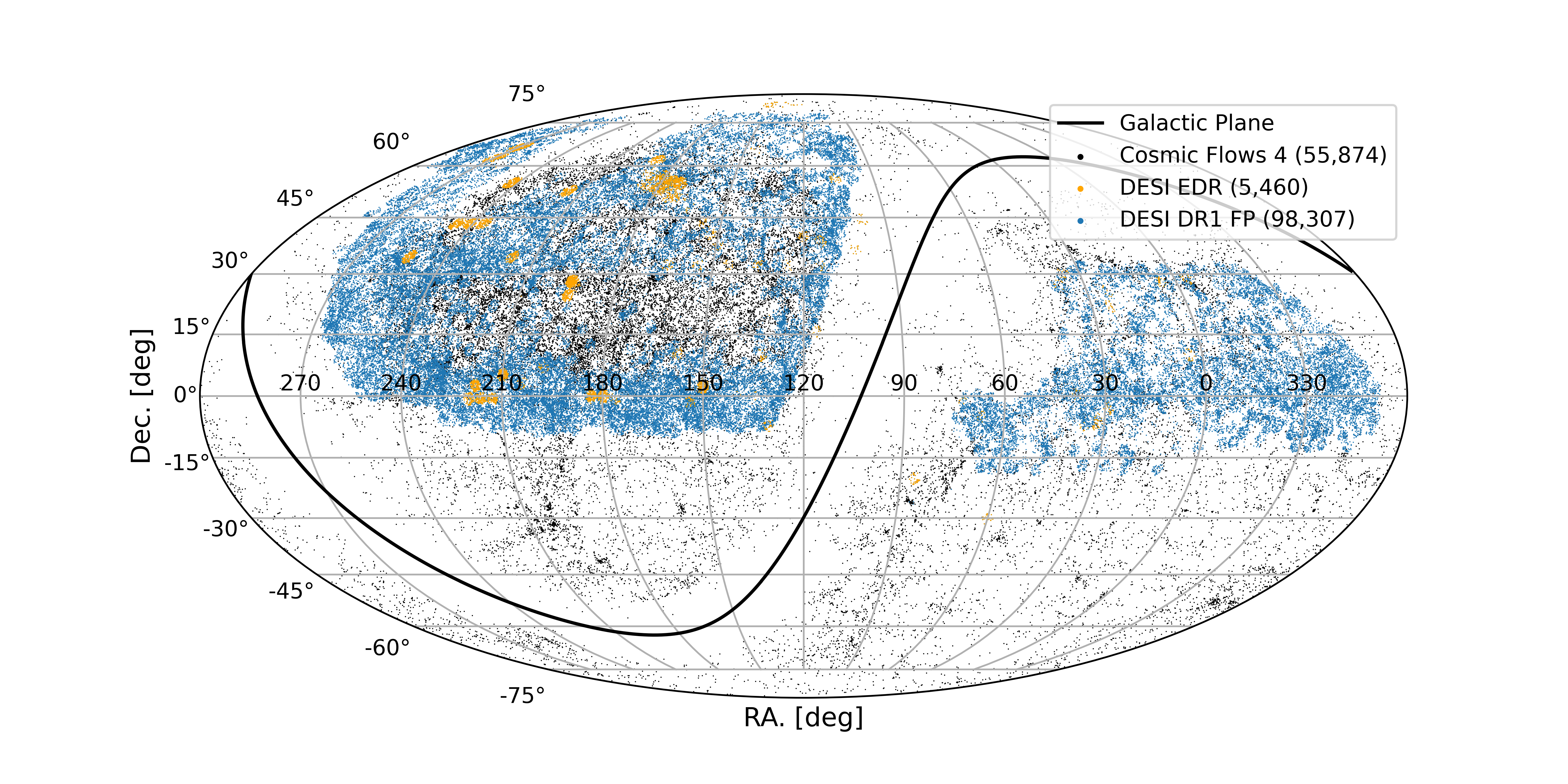}
 \caption{Distribution of the DESI DR1 Fundamental Plane sample (blue) in Right Ascension and Declination compared to the CosmicFlows-4 catalogue \citep[black,][]{cosmicflows-4} and the DESI EDR sample \red{\citep[orange,][]{Said_EDR, Douglass_2025}.}}
 \label{fig:skymap_all_obs}
\end{figure*}

Figure \ref{fig:skymap_all_obs} shows the sky position of each galaxy in our sample (blue), the Early Data Release subsample (orange) and the CosmicFlows-4 catalogue (black). From this plot it can be seen that this work substantially increases the total number of galaxies with measured peculiar velocities when compared to preexisting catalogues. The distribution of these galaxies is much denser than previous surveys --- as expected given the scale and observation rate of the DESI survey --- and closely follows the DESI DR1 bright survey completion map \citep{DESI_collab_DR1}. In the coming data releases, the patchiness of this distribution in the DESI footprint will fill in as the DESI survey nears completion.

\begin{figure}[pbt!]
 \centering
 \includegraphics[width=\linewidth]{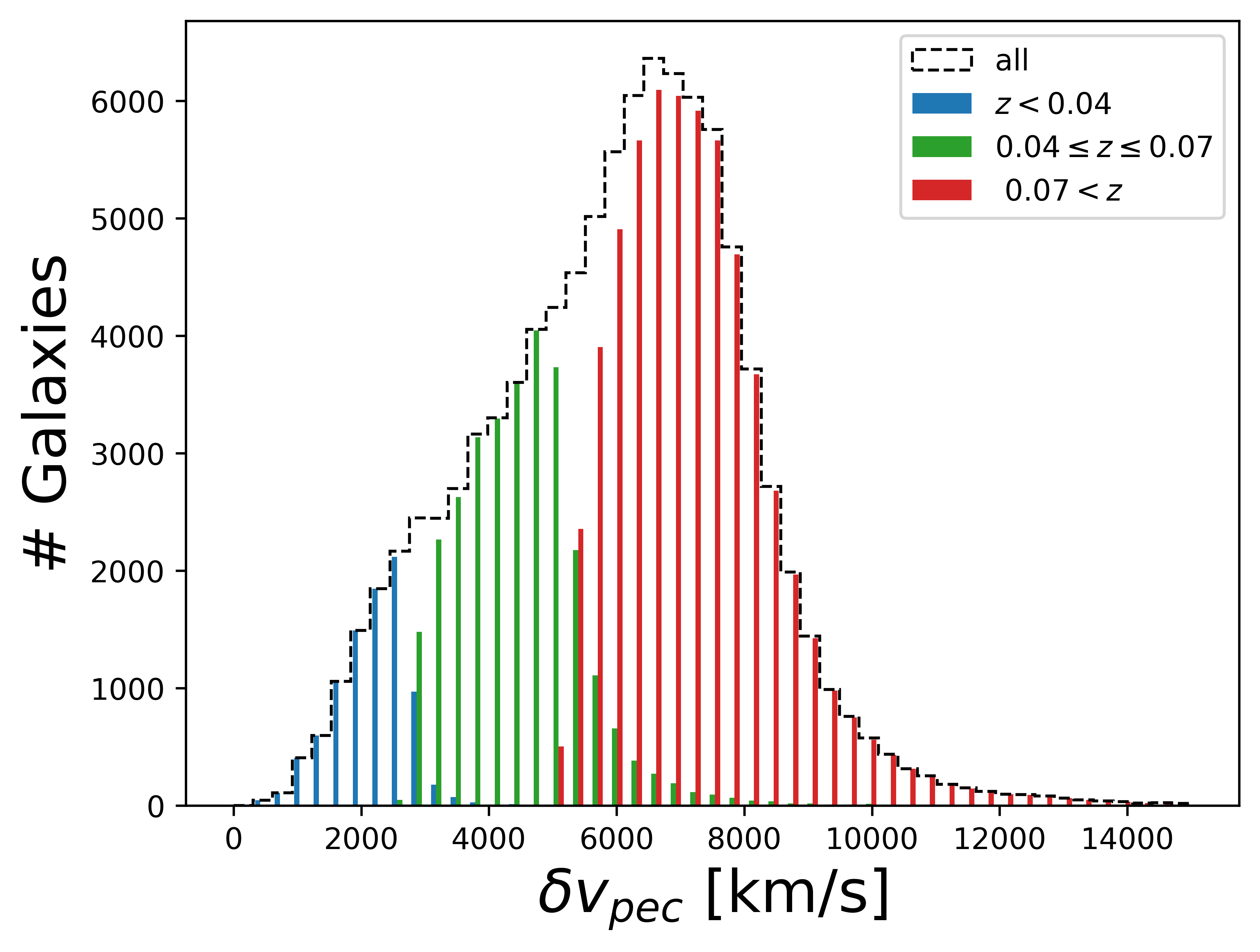}
 \caption{Histogram of the peculiar velocity errors (km/s) for our sample. We also include the breakdown of this distribution into 3 redshift bins: $z<0.4$ (blue), $0.04\leq z\leq 0.07$ (green) and $0.07<z$ (red). This breakdown illustrates how the peculiar velocity errors increase with redshift.\\\\}
 \label{fig:pverrhist}
\end{figure}

We plot the distribution of peculiar velocity errors across our sample in Figure \ref{fig:pverrhist}. We include the distribution over the entire sample (black dashed) as well as in redshift bins, demonstrating how this trend increases with redshift. The uncertainties on individual measurements are large but at $26\%$ of the distance are comparable with previous surveys. The large uncertainty on Fundamental Plane peculiar velocities is the reason it is so important to curate such large catalogues for measuring cosmological parameters.

\begin{figure}[pbt!]
 \centering
 \includegraphics[width=\linewidth]{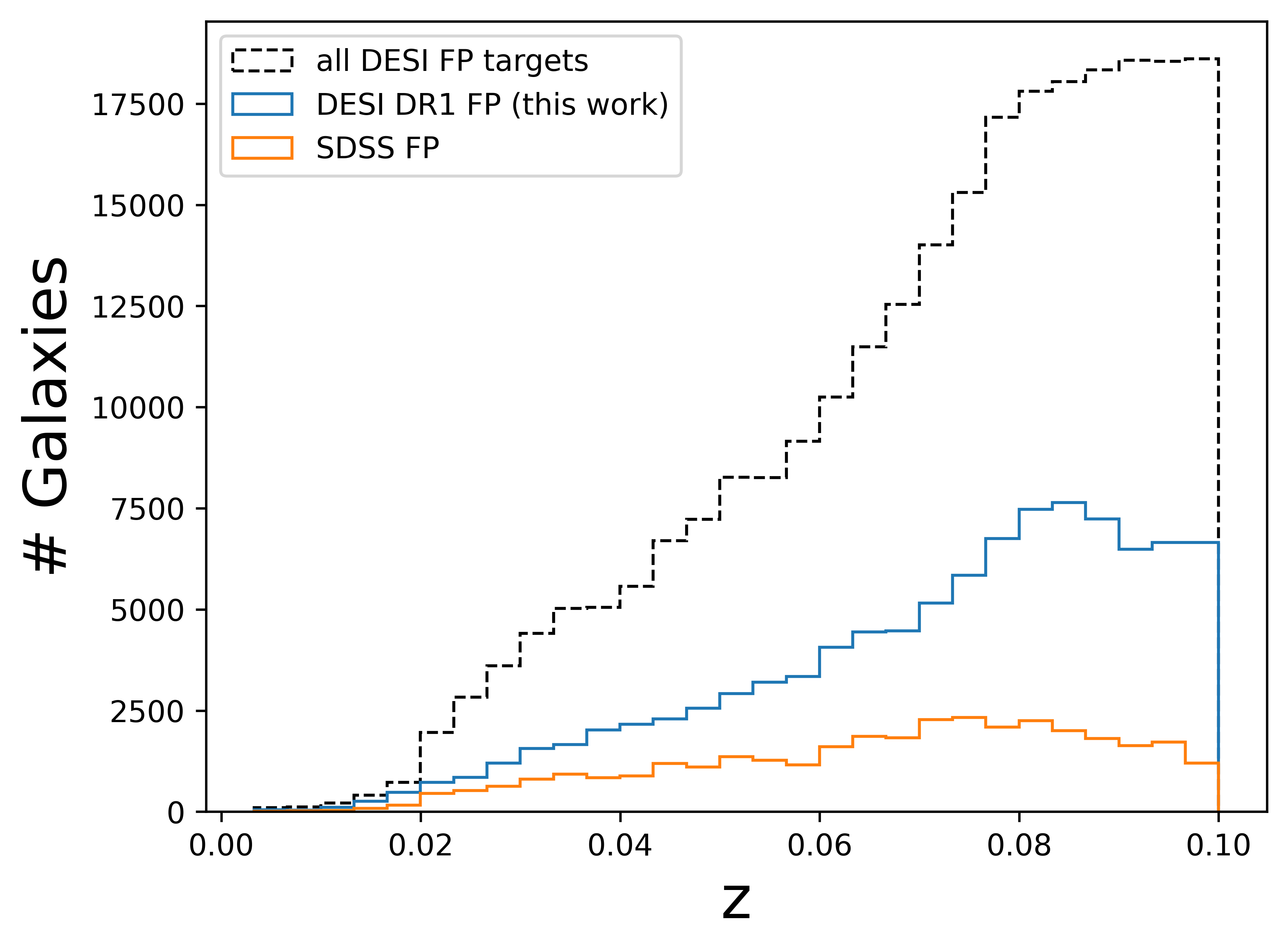}
 \caption{Histogram of the observed redshift of our sample (blue) compared to the full SDSS Fundamental Plane (orange) and compared all potential Fundamental Plane galaxies targeted by DESI over the full survey (black dashed). \red{The DR1 FP sample contains peculiar velocity measurements for $42\%$ of targets that will be observed in the full survey but have already reached $74\%$ of the projected successful sample.}}
 \label{fig:zhist}
\end{figure}

In Figure \ref{fig:zhist}, we show the redshift distribution of our sample (blue) compared to SDSS (orange) and all FP targets that will be observed by DESI during the full survey (dashed black). From this plot it can be seen that in particular our sample supplies a large increase in the number of Fundamental Plane distances at higher redshifts.

\begin{figure*}[hbt!]
    \centering
    \includegraphics[width=\linewidth]{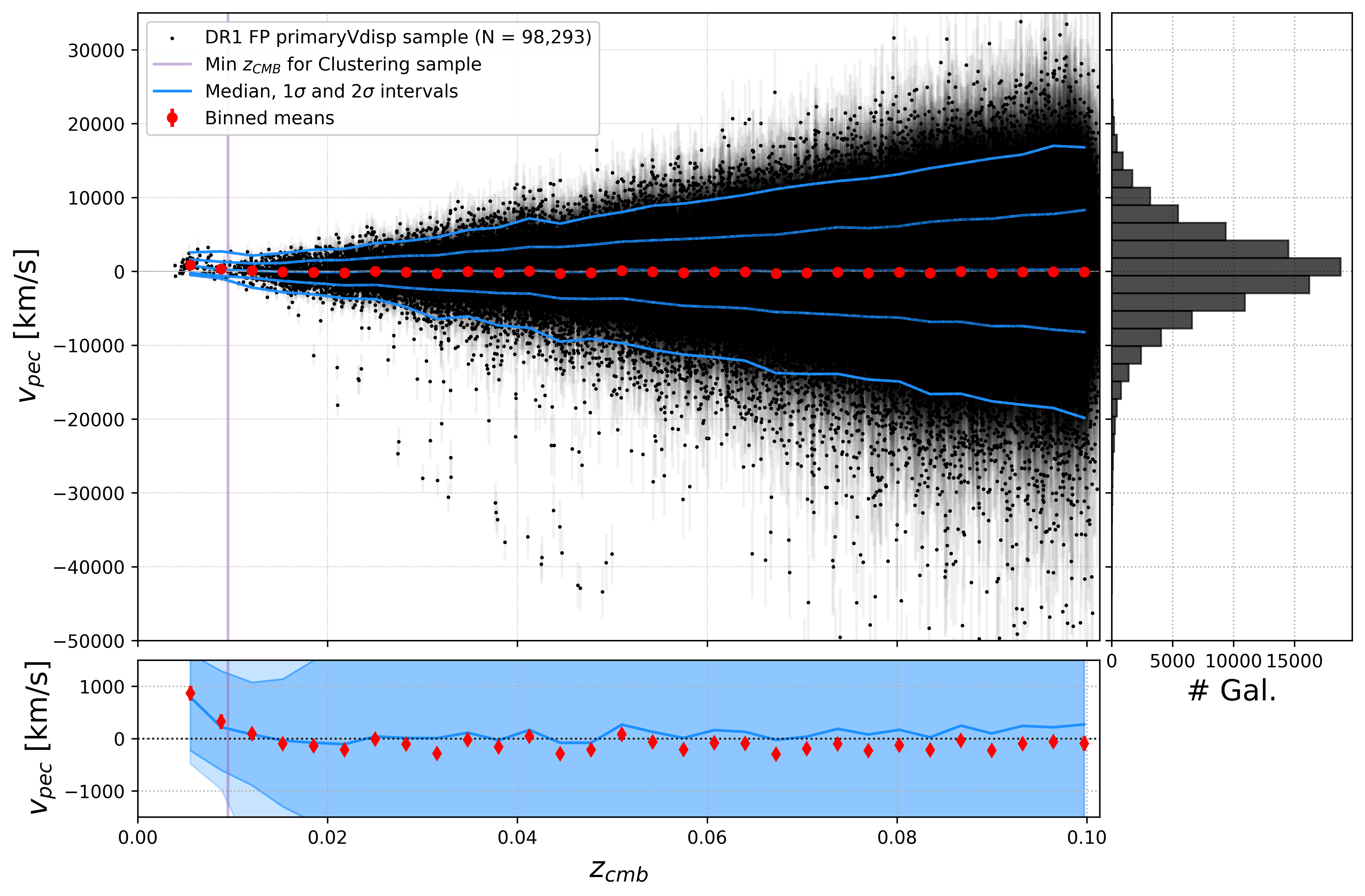}
    \caption{\red{Distribution of measured line-of-sight peculiar velocities as a function of CMB-frame redshift for the DR1 FP \texttt{primaryVdisp} sample (this work). Over-plotted are the $z>0.01$ cut applied to the clustering subsample (purple), together with the  mean peculiar velocities (red) along with the median, $1\sigma$ and $2\sigma$ confidence intervals (blue), all determined in redshift bins. The bottom panel shows a zoomed view of the peculiar velocity axis. Above the clustering sample cut, the binned means fluctuate slightly below $v_\mathrm{pec} =0\kms$ (sample mean and median values $\sim-100\kms$), this is consistent with expectations from variations in large scale structure and the presence of a bulk flow.}}
    \label{fig:vpec_vs_zcmb}
\end{figure*}

\red{Finally, Figure \ref{fig:vpec_vs_zcmb} illustrates the peculiar velocity distribution as a function of CMB-frame redshift. In red we show the mean PV for 30 equally spaced redshift bins. In the zoomed view of the bottom panel it can be seen that these means fluctuate around a flat line slightly below $v_\mathrm{pec}=0\kms$. Being constant with redshift indicates that there is no accelerating or decelerating outflow in the sample. This means that the $H_0$ value fit to the combined DR1 TF and FP data in \citet{DESI_DR1_PV_zp} is a good fit to the FP sample alone. The slight offset from the $v_\mathrm{pec} =0$-axis is consistent with expectations for an outwards bulk flow of $\sim100\kms$, and the fluctuations around this value are what would be expected due to variations in large scale structure. }

\red{
It is only in the nearest two redshift bins that the means differ from expectations. With means at $900\pm 100\kms$ and $300\pm 100\kms$ respectively these values are (borderline) outside the physically explainable range of $\sim250\kms$. This could due to many reasons including: 
\begin{itemize}
    \item the correlation between $v_\mathrm{pec}$ and $z_\mathrm{CMB}$ causing diagonal scattering in this plot;
    \item inherent large-scale structure at low redshift coupled with the incomplete coverage of DESI DR1;
    \item an issue with the $S_n$ or $f_n$ corrections applied during the FP calibration, which more heavily impact low-redshift galaxies;
    \item bias from the peculiar velocity estimator, which is known to occur at low redshift but is usually smaller \citep{Carreres_2023_pv_est};
    \item correlations between our galaxy's velocity and those of nearby galaxies;
    \item and/or a bias in the photometry of nearby galaxies, potentially in the light-profile fit, which would modify the FP inputs for nearby galaxies.
\end{itemize}
This will be further investigated in DR2 where we have a larger sample with a consistent footprint across the sky. 

Overall, Figure \ref{fig:vpec_vs_zcmb} shows little evidence for systematic biases, with peculiar velocity variations consistent with large-scale structure and bulk flows. This is particularly true when the sample is restricted to galaxies with $z>0.01$, as is done in the clustering subsample (Section \ref{section:clustering_cat}).}

\subsection{Clustering Catalogues}\label{section:clustering_cat}

From the full FP catalogue, we construct a subset of galaxy peculiar velocities for the purposes of cosmological measurements, in particular, clustering measurements of positions and velocities for constraining the growth rate of structure as in \citet{DESI_DR1_PV_power_spec, DESI_DR1_PV_maxlike, DESI_DR1_PV_dens_vel_corr}. The clustering catalogue is designed to retain only the most robust peculiar velocity measurements and provide additional systematics weights to account for the effects of redshift completeness, fibre assignment and density variations in the imaging used for target selection.

The FP clustering catalogues are produced by first cutting out outlying galaxies with \red{\texttt{FPCalibrator = True}}, \red{as well as very low redshift galaxies by setting $z_{min}=0.01$.} We then crossmatch the remaining FP galaxies to the BGS large-scale structure (LSS) catalogues presented in \cite{DESI_LSS}, based on their \texttt{targetid}. This keeps only those objects within the well-defined angular mask of the LSS catalogues, and provides the previously mentioned systematics weights for each FP galaxy.

The final FP clustering catalogue contains 73,822 peculiar velocity measurements out of the original set of 98,292 non-duplicate measurements. Two catalogues containing sets of matching random non-clustered data points are also produced, with $20\times$ and $200\times$ the number of real galaxies. These random catalogues have the same sky, redshift and peculiar velocity error distributions. From the random catalogues, we compute a gridded number density of BGS redshifts and FP-based peculiar velocities from which we extrapolate the number density at the location of each real FP galaxy in the clustering catalogue. These are used, e.g., for measurements of the density and velocity correlation functions in \citet{DESI_DR1_PV_dens_vel_corr}. The clustering catalogue and associated random catalogues will be released alongside the full FP sample.

\section{Conclusions}\label{section:conclusion}

In this paper, we present the largest sample of Fundamental Plane distances and peculiar velocities measured to date, \red{with measurements for $98,282$ galaxies. This sample increases the total number of known Fundamental Plane distances by a factor of 2.4 compared to previous works. This represents just the first year of data taken by the DESI survey, which is currently at $42\%$ complete, and as such will increase significantly over the remainder of the survey. During target selection it was projected that DESI would obtain peculiar velocities for $133,000$ Fundamental Plane and $53,000$ Tully-Fisher galaxies \citep{DESI_PV_target_selection}. With the current success rate, the DESI Fundamental Plane survey is on track to exceed this number by $\sim 100,000$ Fundamental Plane galaxies. Such a large peculiar velocity catalogue, particularly one calibrated from a single survey, will be transformative for the field of peculiar velocities and cosmic cartography at low-redshift.

Accompanying this paper we identify several subsamples in our data. These include:
\begin{itemize}
    \item the full catalogue (including duplicate measurements): \\$108,810$ PV measurements;
    \item limiting to a single measurement per galaxy (\texttt{vdispPr-} \texttt{imary = True}): $98,282$ galaxies;
    \item the Fundamental Plane calibration subsample (\texttt{FPcalibr-} \texttt{ator = True}, removes $p<0.01$ outliers): $96,758$ galaxies;
    \item the clustering subsample (used by accompanying works for $H_0$ and $f\sigma_8$ measurements): $73,822$ galaxies.
\end{itemize}
For the majority of use cases we recommend using the \texttt{primar-} \texttt{yVdisp} catalogue as this presents the best peculiar velocity and distance measurement for each galaxy in our catalogue. We believe that sample represents the total number of `good' measurements. This is because further restricting to \texttt{FPcalibrator} removes just $1.6\%$ of the sample, barely above the expected $1\%$ from the $p$-value cut.} 

In a series of companion papers, we present DESI Tully-Fisher distance measurements \citep{DESI_DR1_TF}, simulations and mock data \citep{DESI_DR1_PV_mocks}, and use these data to measure cosmological parameters such as $H_0$ \citep{DESI_DR1_PV_zp} and the growth rate of structure \citep{DESI_DR1_PV_power_spec,DESI_DR1_PV_maxlike,DESI_DR1_PV_dens_vel_corr}. \red{These works (where applicable) use the clustering subsample. This subsample primarily cuts to the DESI Bright Galaxy Survey footprint and is included as the accompanying works require completion accurate clustering weightings and survey footprint masks were previously calculated as part of the DESI DR1 BAO efforts \citep{DESIBAO_2025}.}

\begin{acknowledgement}
This material is based upon work supported by the U.S. Department of Energy (DOE), Office of Science, Office of High-Energy Physics, under Contract No. DE–AC02–05CH11231, and by the National Energy Research Scientific Computing Center, a DOE Office of Science User Facility under the same contract. Additional support for DESI was provided by the U.S. National Science Foundation (NSF), Division of Astronomical Sciences under Contract No. AST-0950945 to the NSF’s National Optical-Infrared Astronomy Research Laboratory; the Science and Technology Facilities Council of the United Kingdom; the Gordon and Betty Moore Foundation; the Heising-Simons Foundation; the French Alternative Energies and Atomic Energy Commission (CEA); the National Council of Humanities, Science and Technology of Mexico (CONAHCYT); the Ministry of Science, Innovation and Universities of Spain (MICIU/AEI/10.13039/501100011033), and by the DESI Member Institutions: \url{https://www.desi.lbl.gov/collaborating-institutions}.

The DESI Legacy Imaging Surveys consist of three individual and complementary projects: the Dark Energy Camera Legacy Survey (DECaLS), the Beijing-Arizona Sky Survey (BASS), and the Mayall z-band Legacy Survey (MzLS). DECaLS, BASS and MzLS together include data obtained, respectively, at the Blanco telescope, Cerro Tololo Inter-American Observatory, NSF’s NOIRLab; the Bok telescope, Steward Observatory, University of Arizona; and the Mayall telescope, Kitt Peak National Observatory, NOIRLab. NOIRLab is operated by the Association of Universities for Research in Astronomy (AURA) under a cooperative agreement with the National Science Foundation. Pipeline processing and analyses of the data were supported by NOIRLab and the Lawrence Berkeley National Laboratory. Legacy Surveys also uses data products from the Near-Earth Object Wide-field Infrared Survey Explorer (NEOWISE), a project of the Jet Propulsion Laboratory/California Institute of Technology, funded by the National Aeronautics and Space Administration. Legacy Surveys was supported by: the Director, Office of Science, Office of High Energy Physics of the U.S. Department of Energy; the National Energy Research Scientific Computing Center, a DOE Office of Science User Facility; the U.S. National Science Foundation, Division of Astronomical Sciences; the National Astronomical Observatories of China, the Chinese Academy of Sciences and the Chinese National Natural Science Foundation. LBNL is managed by the Regents of the University of California under contract to the U.S. Department of Energy. The complete acknowledgements can be found at \url{https://www.legacysurvey.org/}.

Any opinions, findings, and conclusions or recommendations expressed in this material are those of the author(s) and do not necessarily reflect the views of the U. S. National Science Foundation, the U. S. Department of Energy, or any of the listed funding agencies.

The authors are honoured to be permitted to conduct scientific research on I'oligam Du'ag (Kitt Peak), a mountain with particular significance to the Tohono O’odham Nation.
\end{acknowledgement}

\paragraph{Funding Statement}

CER, CH, KS, TMD, acknowledge support from the Australian Government through the Australian Research Council's Discovery Projects funding scheme (project DP250102917).

\paragraph{Competing Interests}

None.

\paragraph{Data Availability Statement}

The DESI DR1 Fundamental Plane catalogue and associated data products \red{from the DESI DR1 PV Survey will be made available upon acceptance of all papers in the series. Once released, they will appear alongside the other DR1 DESI Value-added Catalogues: \url{https://data.desi.lbl.gov/doc/releases/dr1/\#value-added-catalogs}.} Other data products required to reproduce the plots in this paper are available on on Zenodo: \url{https://doi.org/10.5281/zenodo.17784593}. These data products will include a detailed description of the included data files, their respective columns, as well as code to reproduce all figures presented in this work. The columns in our data products include: redshifts, sky locations, photometry, velocity dispersion parameters, fundamental plane parameters, log distance ratios, and the peculiar velocities we derive from these as well as associated errors.

\printendnotes

\bibliography{references}

\appendix

\section*{Affiliations}
{\footnotesize
\textsuperscript{1}School of Mathematics and Physics, The University of Queensland, Brisbane, 4072, Australia\\
\textsuperscript{2}Centre for Extragalactic Astronomy, Durham University, Durham DH1 3LE, United Kingdom\\
\textsuperscript{3}Lawrence Berkeley National Laboratory, 1 Cyclotron Road, Berkeley, CA 94720, USA\\
\textsuperscript{4}Department of Physics, Boston University, 590 Commonwealth Avenue, Boston, MA 02215 USA\\
\textsuperscript{5}Department of Physics, Carnegie Mellon University, 5000 Forbes Avenue, Pittsburgh, PA 15213, USA\\
\textsuperscript{6}Aix Marseille Univ, CNRS/IN2P3, CPPM, Marseille, France\\
\textsuperscript{7}Department of Physics \& Astronomy, University of Rochester, 206 Bausch and Lomb Hall, P.O. Box 270171, Rochester, NY 14627-0171, USA\\
\textsuperscript{8}Dipartimento di Fisica ``Aldo Pontremoli'', Universit\`a degli Studi di Milano, Via Celoria 16, I-20133 Milano, Italy\\
\textsuperscript{9}INAF-Osservatorio Astronomico di Brera, Via Brera 28, 20122 Milano, Italy\\
\textsuperscript{10}Centre for Astrophysics \& Supercomputing, Swinburne University of Technology, P.O. Box 218, Hawthorn, VIC 3122, Australia\\
\textsuperscript{11}Department of Physics \& Astronomy, University College London, Gower Street, London, WC1E 6BT, UK\\
\textsuperscript{12}Korea Astronomy and Space Science Institute, 776, Daedeokdae-ro, Yuseong-gu, Daejeon 34055, Republic of Korea\\
\textsuperscript{13}Instituto de F\'{\i}sica, Universidad Nacional Aut\'{o}noma de M\'{e}xico, Circuito de la Investigaci\'{o}n Cient\'{\i}fica, Ciudad Universitaria, Cd. de M\'{e}xico C.~P.~04510, M\'{e}xico\\
\textsuperscript{14}Department of Astronomy \& Astrophysics, University of Toronto, Toronto, ON M5S 3H4, Canada\\
\textsuperscript{15}Department of Physics \& Astronomy and Pittsburgh Particle Physics, Astrophysics, and Cosmology Center (PITT PACC), University of Pittsburgh, 3941 O'Hara Street, Pittsburgh, PA 15260, USA\\
\textsuperscript{16}University of California, Berkeley, 110 Sproul Hall \#5800 Berkeley, CA 94720, USA\\
\textsuperscript{17}Institut de F\'{i}sica d'Altes Energies (IFAE), The Barcelona Institute of Science and Technology, Edifici Cn, Campus UAB, 08193, Bellaterra (Barcelona), Spain\\
\textsuperscript{18}Departamento de F\'isica, Universidad de los Andes, Cra. 1 No. 18A-10, Edificio Ip, CP 111711, Bogot\'a, Colombia\\
\textsuperscript{19}Observatorio Astron\'omico, Universidad de los Andes, Cra. 1 No. 18A-10, Edificio H, CP 111711 Bogot\'a, Colombia\\
\textsuperscript{20}Institut d'Estudis Espacials de Catalunya (IEEC), c/ Esteve Terradas 1, Edifici RDIT, Campus PMT-UPC, 08860 Castelldefels, Spain\\
\textsuperscript{21}Institute of Cosmology and Gravitation, University of Portsmouth, Dennis Sciama Building, Portsmouth, PO1 3FX, UK\\
\textsuperscript{22}Institute of Space Sciences, ICE-CSIC, Campus UAB, Carrer de Can Magrans s/n, 08913 Bellaterra, Barcelona, Spain\\
\textsuperscript{23}University of Virginia, Department of Astronomy, Charlottesville, VA 22904, USA\\
\textsuperscript{24}Fermi National Accelerator Laboratory, PO Box 500, Batavia, IL 60510, USA\\
\textsuperscript{25}Center for Cosmology and AstroParticle Physics, The Ohio State University, 191 West Woodruff Avenue, Columbus, OH 43210, USA\\
\textsuperscript{26}Department of Physics, The Ohio State University, 191 West Woodruff Avenue, Columbus, OH 43210, USA\\
\textsuperscript{27}The Ohio State University, Columbus, 43210 OH, USA\\
\textsuperscript{28}Department of Physics, University of Michigan, 450 Church Street, Ann Arbor, MI 48109, USA\\
\textsuperscript{29}University of Michigan, 500 S. State Street, Ann Arbor, MI 48109, USA\\
\textsuperscript{30}Department of Physics, The University of Texas at Dallas, 800 W. Campbell Rd., Richardson, TX 75080, USA\\
\textsuperscript{31}NSF NOIRLab, 950 N. Cherry Ave., Tucson, AZ 85719, USA\\
\textsuperscript{32}Sorbonne Universit\'{e}, CNRS/IN2P3, Laboratoire de Physique Nucl\'{e}aire et de Hautes Energies (LPNHE), FR-75005 Paris, France\\
\textsuperscript{33}Department of Astronomy and Astrophysics, UCO/Lick Observatory, University of California, 1156 High Street, Santa Cruz, CA 95064, USA\\
\textsuperscript{34}Department of Astronomy and Astrophysics, University of California, Santa Cruz, 1156 High Street, Santa Cruz, CA 95065, USA\\
\textsuperscript{35}Department of Astronomy, The Ohio State University, 4055 McPherson Laboratory, 140 W 18th Avenue, Columbus, OH 43210, USA\\
\textsuperscript{36}Instituci\'{o} Catalana de Recerca i Estudis Avan\c{c}ats, Passeig de Llu\'{\i}s Companys, 23, 08010 Barcelona, Spain\\
\textsuperscript{37}Department of Physics and Astronomy, Siena University, 515 Loudon Road, Loudonville, NY 12211, USA\\
\textsuperscript{38}IRFU, CEA, Universit\'{e} Paris-Saclay, F-91191 Gif-sur-Yvette, France\\
\textsuperscript{39}Department of Physics and Astronomy, University of Waterloo, 200 University Ave W, Waterloo, ON N2L 3G1, Canada\\
\textsuperscript{40}Perimeter Institute for Theoretical Physics, 31 Caroline St. North, Waterloo, ON N2L 2Y5, Canada\\
\textsuperscript{41}Waterloo Centre for Astrophysics, University of Waterloo, 200 University Ave W, Waterloo, ON N2L 3G1, Canada\\
\textsuperscript{42}Space Sciences Laboratory, University of California, Berkeley, 7 Gauss Way, Berkeley, CA 94720, USA\\
\textsuperscript{43}Instituto de Astrof\'{i}sica de Andaluc\'{i}a (CSIC), Glorieta de la Astronom\'{i}a, s/n, E-18008 Granada, Spain\\
\textsuperscript{44}Departament de F\'isica, EEBE, Universitat Polit\`ecnica de Catalunya, c/Eduard Maristany 10, 08930 Barcelona, Spain\\
\textsuperscript{45}Department of Physics and Astronomy, Sejong University, 209 Neungdong-ro, Gwangjin-gu, Seoul 05006, Republic of Korea\\
\textsuperscript{46}CIEMAT, Avenida Complutense 40, E-28040 Madrid, Spain\\
\textsuperscript{47}National Astronomical Observatories, Chinese Academy of Sciences, A20 Datun Road, Chaoyang District, Beijing, 100101, P.~R.~China
}

\end{document}